\documentclass[%
 reprint,
showpacs,preprintnumbers,
 amsmath,amssymb,
 aps,
]{aastex61}

\usepackage{amsmath}
\usepackage{graphicx}
\usepackage{dcolumn}
\usepackage{bm}
\usepackage{multirow}
\usepackage{float}

\shorttitle{Nuclear Neutrino Spectra}
\shortauthors{Misch et al.}

\begin{document}

\title{Neutrino Spectra from Nuclear Weak Interactions in $sd$-Shell Nuclei Under Astrophysical Conditions}

\author{G. Wendell Misch}
\affiliation{School of Physics and Astronomy \\
Shanghai Jiao Tong University \\
Shanghai 200240, China}
\affiliation{Collaborative Innovation Center of IFSA \\
Shanghai Jiao Tong University \\
Shanghai 200240, China}
\email{wendell@sjtu.edu.cn, wendell.misch@gmail.com (preferred)}

\author{Yang Sun}
\affiliation{School of Physics and Astronomy \\
Shanghai Jiao Tong University \\
Shanghai 200240, China}
\affiliation{Collaborative Innovation Center of IFSA \\
Shanghai Jiao Tong University \\
Shanghai 200240, China}
\affiliation{Institute of Modern Physics \\
Chinese Academy of Sciences \\
Lanzhou 730000, China}
\email{sunyang@sjtu.edu.cn}


\author{George M. Fuller}
\affiliation{Department of Physics \\
University of California, San Diego \\
La Jolla, CA 92093, USA}
\email{gfuller@ucsd.edu}

\date{\today}

\begin{abstract}
We present shell model calculations of nuclear neutrino energy spectra for 70 $sd$-shell nuclei over the mass number range $A=21-35$.  Our calculations include nuclear excited states as appropriate for the hot and dense conditions characteristic of pre-collapse massive stars.  We consider neutrinos produced by charged lepton captures and decays and, for the first time in tabular form, neutral current nuclear deexcitation, providing neutrino energy spectra on the Fuller-Fowler-Newman temperature-density grid for these interaction channels for each nucleus.  We use the full $sd$-shell model space to compute initial nuclear states up to 20 MeV excitation with transitions to final states up to 35-40 MeV, employing a modification of the Brink-Axel hypothesis to handle high temperature population factors and the nuclear partition functions.

\end{abstract}


\section{Introduction}
\label{sec:intro}

In this paper we calculate key nuclear interactions that eventually could help develop forensic evidence of the principle agent in the demise of massive stars -- the weak interaction.  We here consider $sd$-shell, intermediate-mass nuclei produced during core and shell carbon/oxygen, neon/magnesium, and the beginning of silicon burning in stars which are the progenitors of core-collapse supernovae.  It has been pointed out that neutrinos from these stars in their pre-collapse phases may be detectable  \citep{omk:2004a,omk:2004b,oh:2010,asakura-etal:2016,yoshida-etal:2016,plf:2017,kato-etal:2017}.  However, the detectability of these neutrinos hinges on the neutrino energy spectra produced by hot nuclei.  In turn, \cite{mf:2016} have shown that nuclear structure, properly accounted for, can enhance the production of the highest-energy, most-detectable neutrinos produced by these hot, excited nuclei.  Here we survey 70 $sd$-shell nuclei that are likely most relevant for pre-supernova neutrino production.  We will address heavier $pf$-shell nuclei in forthcoming work.

The weak interaction dominates the evolution of pre-core collapse stars. The weak interaction conversion of protons into neutrons facilitates the fusion of increasingly massive and neutron-rich nuclear fuels from hydrogen burning on the main sequence to the end of fusion energy sources with the production of iron-peak nuclei in nuclear statistical equilibrium in the pre-collapse ``iron'' core. Of course, the Coulomb barriers for nuclear fusion go up dramatically in latter stages of hydrostatic nuclear evolution, requiring a higher and higher central temperature to enable nuclear reactions \citep{arnett:1977,bbal:1979,langanke:2015}.

However, neutrino emission via plasma neutrino pair bremsstrahlung processes \citep{itoh-etal:1996} and via nuclear weak interactions (e.g., accompanying positron decay and electron capture) \emph{necessarily} accompanies the nuclear evolution of the pre-supernova star \citep{ww:1995,whw:2002}. The result is that these massive stars are in a sense giant \lq\lq refrigerators,\rq\rq\ with neutrino emission removing entropy from the stellar core, and so running down this quantity from an entropy-per-baryon (in units of Boltzmann’s constant $k_{\rm b}$) $s \sim 10$ during core hydrogen burning, to $s \sim 1$ in the iron core at the onset of collapse. While the state of the material in the stellar core may be highly ordered, and therefore thermodynamically \lq\lq cold,\rq\rq\ the temperature can be high. For example, the iron core may have a temperature $T \sim 1\,{\rm MeV}$, albeit with a density approaching $\rho \sim {10}^{10}\,{\rm g}\,{\rm cm}^{-3}$, giving relativistically degenerate conditions for electrons, with Fermi levels $\mu_e \approx 11.1\,{\rm MeV}\,{\left( \rho_{10}\, Y_e \right)}^{1/3}$, with $\rho_{10}$ the density in units of ${10}^{10}\,{\rm g}\,{\rm cm}^{-3}$ and $Y_e$ the electron fraction, the net number of electrons per baryon. Electrons supply the bulk of the pressure support in hydrostatic equilibrium in the iron core. The low entropy, relativistic electron degenerate conditions set the core up for instability. In summary, the integrated history of neutrino emission in the star brings it to this perilous end state. Much of that neutrino emission comes from hot, excited intermediate mass nuclei \citep{bbal:1979,bw:1982,sjfk:2008,liebendorfer-etal:2008,liebendorfer-etal:2009,hix-etal:2010,bbol:2011}.

Calculating the nuclear weak interaction processes that occur in the hydrostatic run-up to this end state is challenging.  The potentially high temperatures and considerable electron Fermi energies characterizing material in the central regions of these stars imply that nuclear excited states contribute to important weak interactions in both neutral \citep{btz:1974,fm:1991} and charged current channels \citep{bahcall:1964}, and electron degeneracy can leverage the energy dependence of weak interaction processes, producing large continuum electron capture rates \citep{ffn:1980,ffn:1982a,ffn:1982b,ffn:1985,oda-etal:1994,lm:1999,lm:2000,lm:2001,ferreira-etal:2014,lm:2014,martinez-pinedo-etal:2014,nsf:2016,sarriguren:2016,rgo:2017}.  Nuclear weak interactions also play important roles in many other extreme astrophysical environments \citep{wh:2015,iliadis-etal:2016,liu-etal:2016,mernier-etal:2016,mori-etal:2016,rn:2016,stn:2016}.  These rates are mediated by nuclear structure.  However, despite tremendous advances in method and technology in modern nuclear laboratories, direct experimental measurement of most nuclear levels and nuclear transition matrix elements is terrestrially impossible.  Therefore, to obtain the weak interaction rates and, as we do in this paper, associated neutrino energy spectra, reliable theoretical calculation is the only choice, taking care that the techniques match experimental results where they are available \citep{noji-etal:2014}.

\begin{figure}
\centering
\includegraphics[scale=.4]{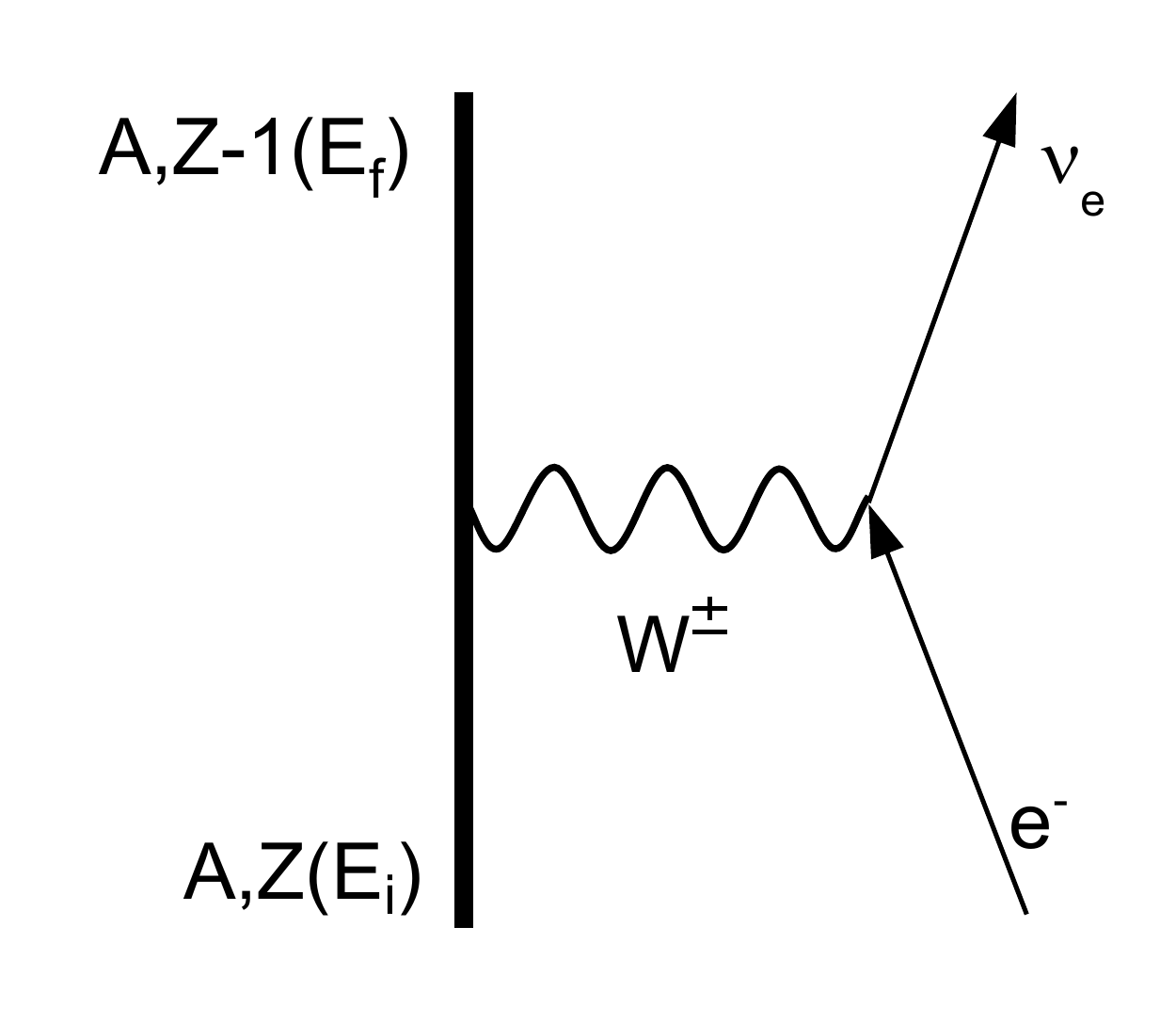}
\includegraphics[scale=.4]{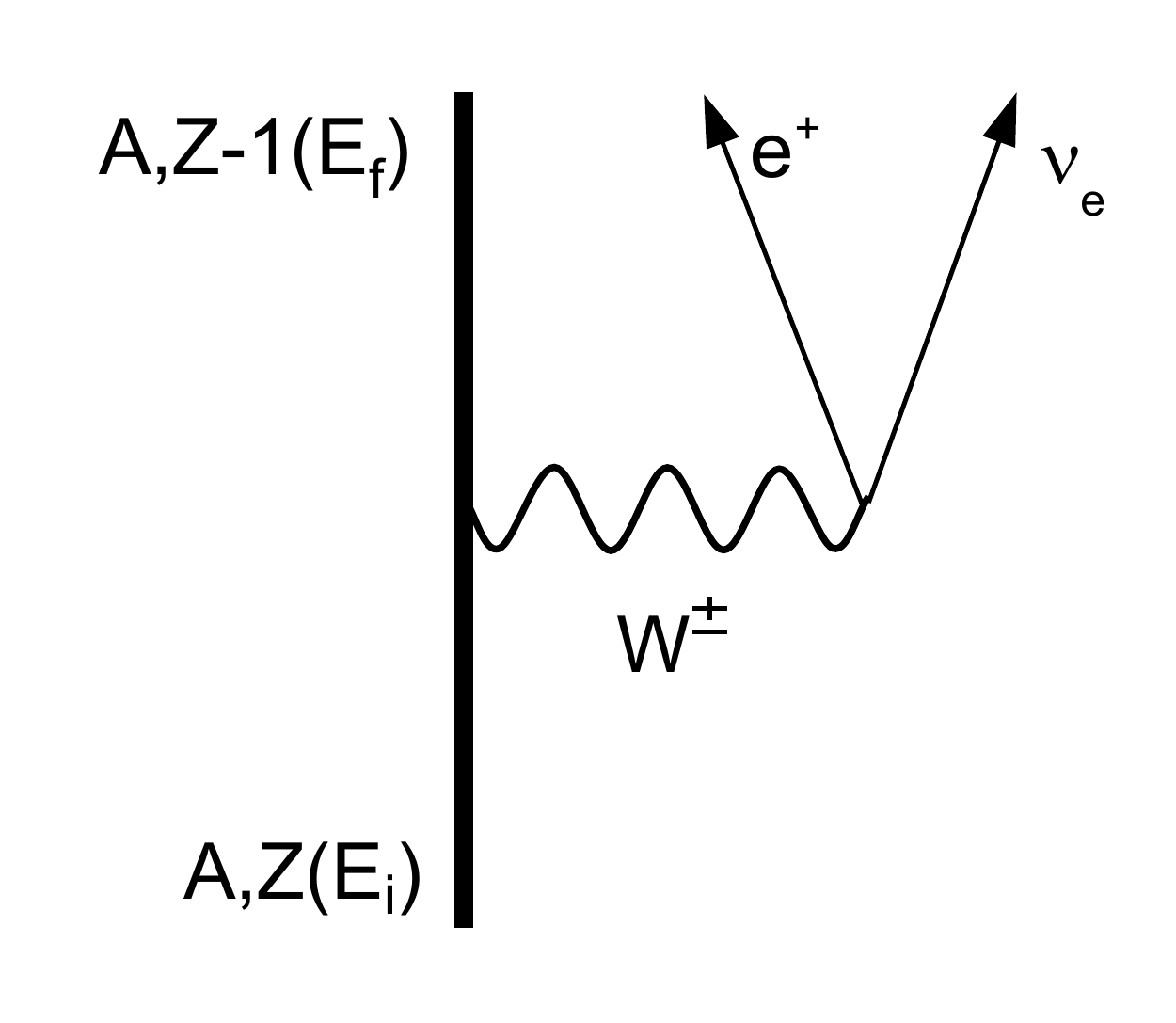}\\
\includegraphics[scale=.4]{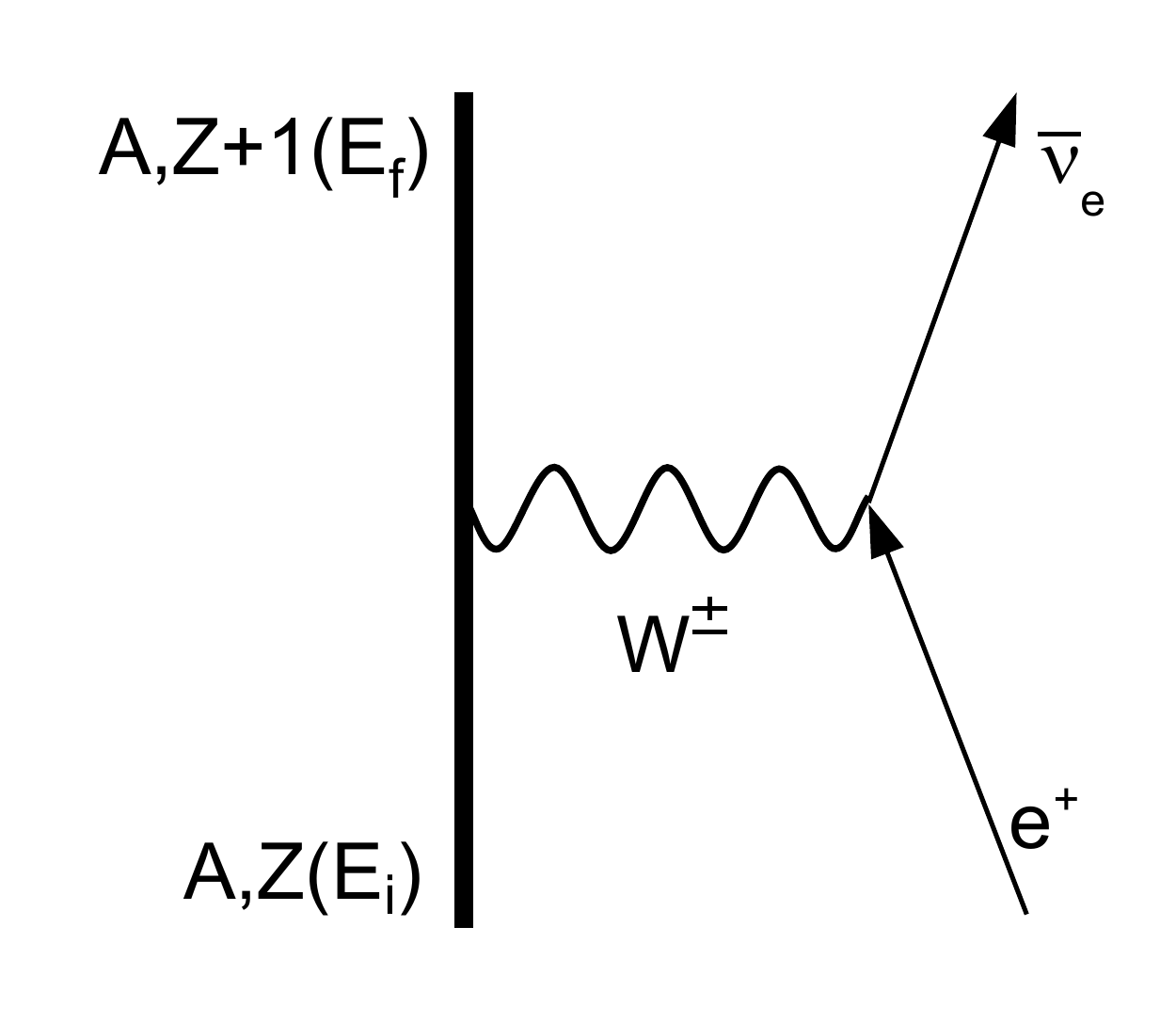}
\includegraphics[scale=.4]{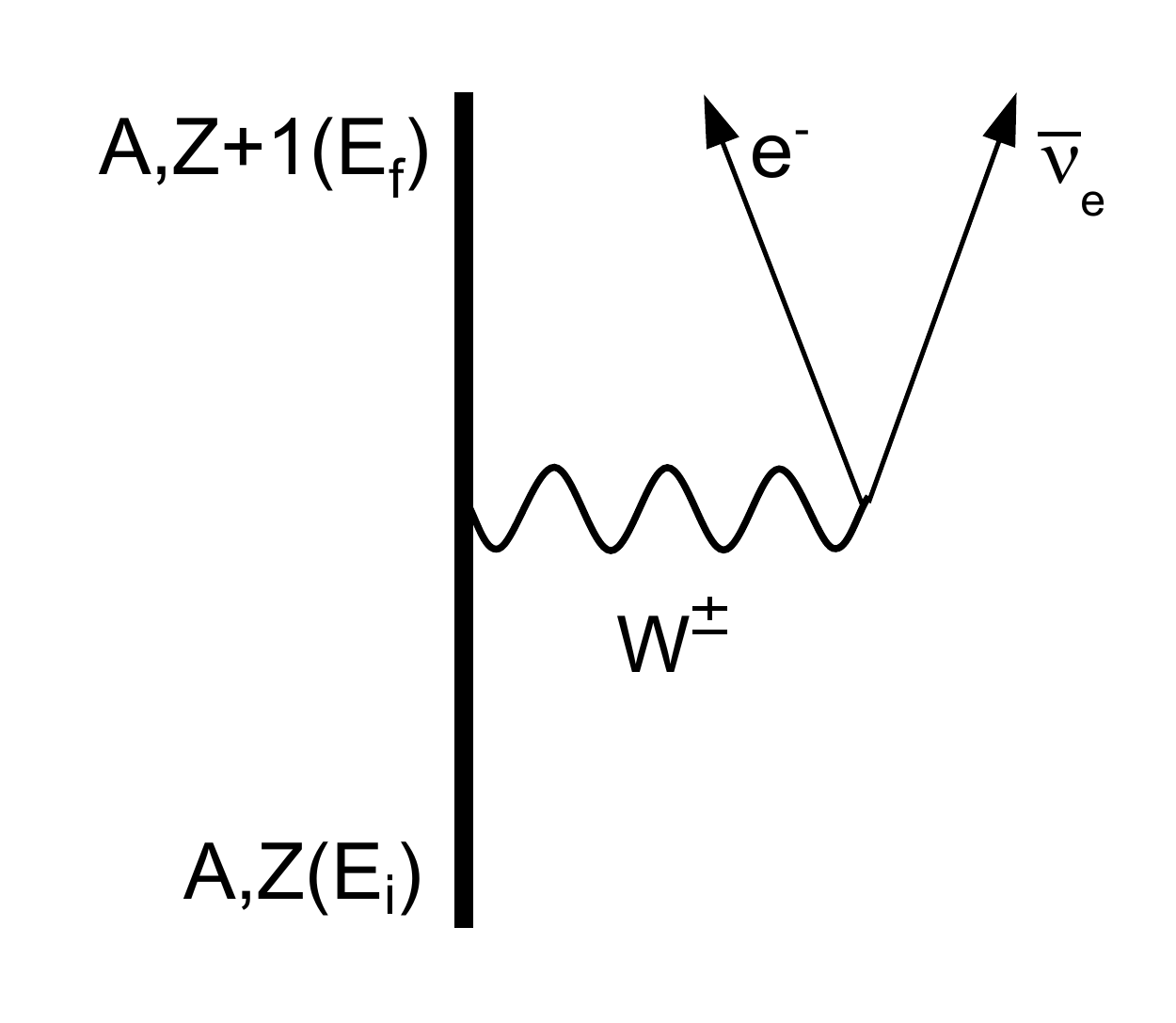}\\
\includegraphics[scale=.4]{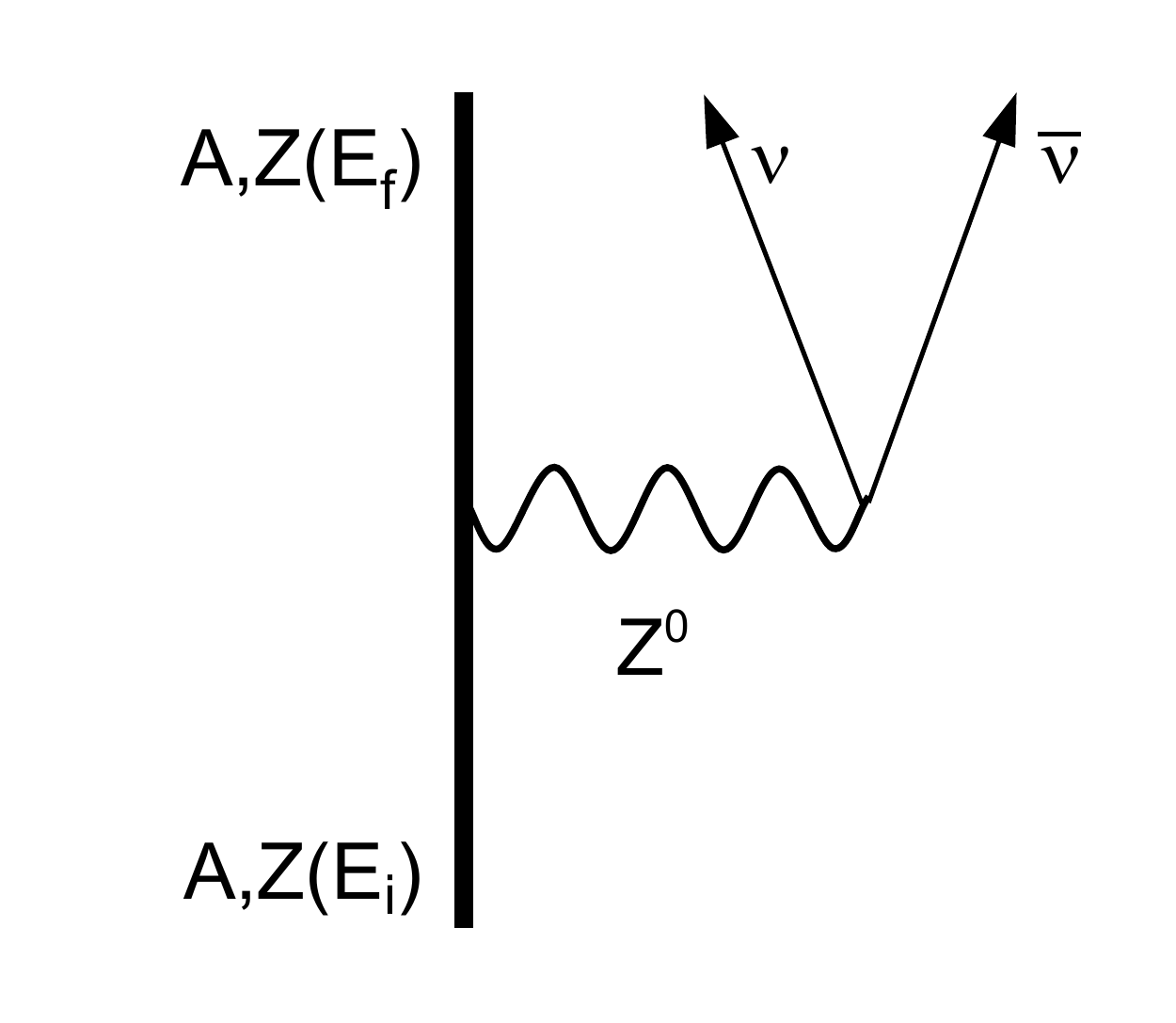}
\caption{The five nuclear weak processes computed in this work.  The upper two, electron capture and positron decay, produce electron flavored neutrinos.  The middle two, positron capture and electron decay, produce electron antineutrinos.  The bottom, neutral current deexcitation, produces neutrino-antineutrino pairs of all flavors.}
\label{fig:feyn}
\end{figure}

Calculation of the rates of these processes is difficult, but because the neutrino emission in pre-collapse stars may be detectable if they are close enough, the stakes may be significant.  Prospects for neutrino detection increase considerably when the neutrino energies are larger, and the highest energy neutrinos may be produced by weak interactions that proceed through thermally populated excited nuclear states; this couples in the vagaries of nuclear structure. This is the subject of this work, in which we compute the neutrino energy spectra for the five reactions shown in figure \ref{fig:feyn}.

In section \ref{sec:calculations} we describe our methods for computing neutrino spectra.  Section \ref{sec:phase} gives the reaction phase space factors, section \ref{sec:nuclear} details our calculations of nuclear structure and how we match our calculations with experiment, and section \ref{sec:spectra} shows how to combine these elements into neutrino energy spectra.  Section \ref{sec:discussion} provides discussion of our results, including notes on how to get the most accurate results from our tabulated spectra.

\section{Calculations}
\label{sec:calculations}

Nuclear neutrino production rates and spectra are calculated from the product of three primary ingredients: physical constants, which are measured in the laboratory; nuclear structure, which includes nuclear energy levels and nuclear transition matrix elements; and phase space factors, which account for the dynamics of the interacting particles.  The general rate of weak reaction $X$ from an initial parent nuclear state $\vert i\rangle$ to a final daughter nuclear state $\vert f\rangle$ takes the form

\begin{equation}
\label{eq:rate_generic}
\lambda_{if}^{X}=\lambda_0^{X}B_{if}^{X}f_{if}^{X},
\end{equation}
where $\lambda_0^X$ contains the relevant physical constants, $B_{if}^X$ is the nuclear transition strength, and $f_{if}^X$ is the phase space factor.

For the reactions considered here, there are three relevant sets of physical constants: two for charged current (CC) nuclear transitions and one for neutral current (NC) transitions.  The two CC sets of constants correspond to the allowed isospin raising/lowering Gamow-Teller operators (GT$_\pm$) and the super allowed isospin raising/lowering Fermi operators (F$_\pm$), each of which can convert a neutron to a proton (raise isospin, increasing nuclear charge) or convert a proton to a neutron (lower isospin, decreasing nuclear charge).  We use the particle physics convention that up quarks are isospin up (have a positive third component of isospin), and down quarks are isospin down; this yields isospin-up protons and isospin-down neutrons.  The NC set of constants is associated with the third component of isospin in the Gamow-Teller operator (GT$_3$), which preserves the nuclear charge.  All of these operators are described further in section \ref{sec:nuclear}.

The phase space factor is specific to the reaction: electron capture (ec), positron decay (pd), positron capture (pc), electron decay (ed), and neutral current deexcitation (de).  The nuclear transition strengths depend on (among other things) the reaction channel: charge increases (GT$_+$, F$_+$), charge decreases (GT$_-$, F$_-$), and charge is unchanged (GT$_3$).  Thus, the electron capture rate from nuclear state $\vert i\rangle$ to state $\vert f\rangle$ has the labels

\begin{equation}
\lambda_{if}^{ec}=\left(\lambda_0^{GT_-}B_{if}^{GT_-}+\lambda_0^{F_-}B_{if}^{F_-}\right)f_{if}^{ec},
\end{equation}
while the deexcitation rate will take the labels

\begin{equation}
\lambda_{if}^{de}=\lambda_0^{GT_3}B_{if}^{GT_3}f_{if}^{de}.
\end{equation}

Because physical constants come directly from experiment and only scale the reaction rates and spectra, we will accept them as provided in our references and not discuss them in depth here.  The other two ingredients (nuclear structure and phase-space factors), however, require careful and specific treatment.  In this section, we describe our approach.

\subsection{Phase Space Factors and Physical Constants}
\label{sec:phase}

In computing rates, the phase space factors take the form of integrals over incoming and outgoing particle momenta (or equivalently, energies).  Following \cite{ffn:1980}, the phase space factor for charged lepton capture on parent nucleus state $\vert i\rangle$ going to daughter nucleus state $\vert f\rangle$ is

\begin{equation}
\label{eq:phase_cap}
f_{if}^{ec,pc}=\int_{max(1,q_{if})}^\infty dw_{e^\mp}~w_{e^\mp}^2(-q_{if}+w_{e^\mp})^2G(Z,w_{e^\mp})f_{e^\mp}(w_{e^\mp})[1-f_{\nu,\overline{\nu}}(-q_{if}+w_{e^\mp})],
\end{equation}
and the phase space factor for charged lepton emission is
\begin{equation}
\label{eq:phase_dec}
f_{if}^{ed,pd}=\int_1^{-q_{if}}dw_{e^\mp}~w_{e^\mp}^2(-q_{if}-w_{e^\mp})^2G(Z,w_{e^\mp})[1-f_{e^\mp}(w_{e^\mp})][1-f_{\overline{\nu},\nu}(-q_{if}-w_{e^\mp})].
\end{equation}

In equations \ref{eq:phase_cap} and \ref{eq:phase_dec}, $w_{e^\mp}$ is the charged lepton energy; $q_{if}$ is the nuclear transition energy, given by $m_f+e_f-(m_i+e_i)$, where $m$ and $e$ are the masses and excitation energies of the initial and final nuclear states; $G(Z,w_{e^\mp})$ is the Coulomb correction to the charged lepton wave function, described in detail in \cite{ffn:1980}, with $Z$ the nuclear charge; and $f_{e^\mp}(w_{e^\mp})$ is the charged lepton occupation probability, taken throughout this work to be a Fermi-Dirac distribution (upper signs for electrons, lower signs for positrons).  The lowercase symbols for energies and masses indicate that they are measured in units of electron mass $m_e$.

In the capture equation, $-q_{if}+w_{e^\mp}$ is the outgoing neutrino energy, and the occupation probability $f_{e^\mp}(w_{e^\mp})$ corresponds to the available charged lepton flux.  The lower limit on the integral indicates that the captured lepton must (of course) have at least as much energy as its rest mass, and furthermore, it must have enough energy for the nucleus to make the transition.  In the emission equation, $-q_{if}-w_{e^\mp}$ is the neutrino energy, $[1-f_{e^\mp}(w_{e^\mp})]$ is the outgoing charged lepton blocking factor, and the integration limits indicate that the outgoing charged lepton must have at least its rest mass energy and at most the total energy lost by the nucleus in the transition.  In both equations, $1-f_{\nu,\overline{\nu}}$ is the outgoing neutrino blocking factor.

From \cite{fm:1991}, the phase space factor for neutral current deexcitation is

\begin{equation}
\label{eq:phase_deex}
f_{if}^{de}=\int_0^{-Q_{if}}dE_\nu~E_\nu^2(-Q_{if}-E_\nu)^2[1-f_\nu(E_\nu)][1-f_{\overline{\nu}}(-Q_{if}-E_\nu)],
\end{equation}
where $Q_{if}$ is the nuclear transition energy, $E_\nu$ is the neutrino energy, and $(-Q_{if}-E_\nu)$ is the antineutrino energy.  The uppercase symbols for energies indicate that the choice of units is arbitrary.

Throughout this work, we present results with MeV as the unit of energy and seconds as the unit of time, which is sufficient to assign numerical values to the physical constants.  In the charged current channels, this choice of units coupled with the specific expressions of equations \ref{eq:phase_cap} and \ref{eq:phase_dec} gives

\begin{equation}
\label{eq:cc_constants}
\begin{aligned}
\lambda_0^{GT^\pm}&=\frac{\rm ln(2)}{D}\left(\frac{g_A}{g_V}\right)^2&&=1.8182\times 10^{-4}~{\rm s}^{-1}\\
\lambda_0^{F^\pm}&=\frac{\rm ln(2)}{D}&&=1.1283\times 10^{-4}~{\rm s}^{-1},
\end{aligned}
\end{equation}
where $D=\frac{2\pi^3\hbar^7}{g_V^2m_e^5c^4}\approx	6143$ s \citep{ht:2009} and $g_A/g_V\approx -1.2694$ is the ratio of the axialvector and vector coupling constants \citep{rpp:2010}.  In the neutral current channel, we have

\begin{equation}
\label{nc_constants}
\lambda_0^{GT_3}=\frac{G_F^2}{2\pi^3\hbar}\left(\frac{g_A}{g_V}\right)^2\approx 5.371\times 10^{-3}~{\rm MeV^{-5}s^{-1}},
\end{equation}
where $G_F\approx 1.166\times 10^{-11}$ MeV$^{-2}$ is the Fermi constant \citep{fm:1991}.

\subsection{Nuclear Structure}
\label{sec:nuclear}

To obtain nuclear energy levels and transition strengths, we used the shell model code OXBASH \citep{oxbash} in the $sd$-shell model space.  The $sd$-shell model space assumes an inert $^{16}$O core whose nucleons always occupy the 16 lowest single-particle energy levels, a valence shell consisting of the $2s$ and $1p$ single particle orbitals, and all higher single particle levels inaccessible.  Because the $1s$ and $2d$ single particle states have positive parity, this will necessarily exclude negative parity states from our calculations, but these states likely do not have a large effect on the neutrino spectra.  In $sd$-shell nuclei, the low-lying states have positive parity, so for most real nuclei, negative parity states do not contribute much to the nuclear partition function; hence, they do not strongly impact the population factors and concomitant neutrino production of positive-parity states.  Neither are negative-parity states as populated as positive-parity states, so we neglect allowed transitions between negative-parity states, trusting them to be comparatively small.

We neglect forbidden transitions of all kinds, including those between positive- and negative-parity states (which involve nuclei crossing major shells, disallowed by our model space) and those which involve spin transitions forbidden by our operators.  Late in collapse, forbidden transitions dominate the electron capture rates on relatively heavy, extremely neutron rich nuclei, because the high neutron Fermi energy in the nucleus blocks allowed electron capture \citep{fuller:1982}, but the nuclei here suffer no such blocking and hence all have much greater allowed (Gamow-Teller) and super allowed (Fermi) strength.

While it is in principle possible to include all of the nuclear states in this model space, it is not computationally feasible to do so.  Therefore, we use the technique of \cite{mf:2016}, which built on the weak strength Brink hypothesis modification of \cite{mfb:2014}.  To whit, we consider all initial states up to 15 MeV individually.  Above 15 MeV, we average together (weighted by spin degeneracy $2J+1$) the lowest 20 states in each of 1 MeV increments up to 20 MeV, giving six high energy average states at each of approximately 15, 16, 17, 18, 19, and 20 MeV.  Each of these high energy average states is assigned the total thermal weight of all states in its bin.  Transition strengths are computed to all final states with energy less than 20 MeV above the highest energy state in each bin.  So, individually considered initial states have strengths computed to final states up to $\sim 35$ MeV, while final states from the 18 MeV bin go up to $\sim 38$ MeV.  This method does not strictly obey detailed balance of thermal strength, but as \cite{misch:2017} showed, that will not meaningfully impact our results.

\begin{figure}
\centering
\includegraphics[scale=.46]{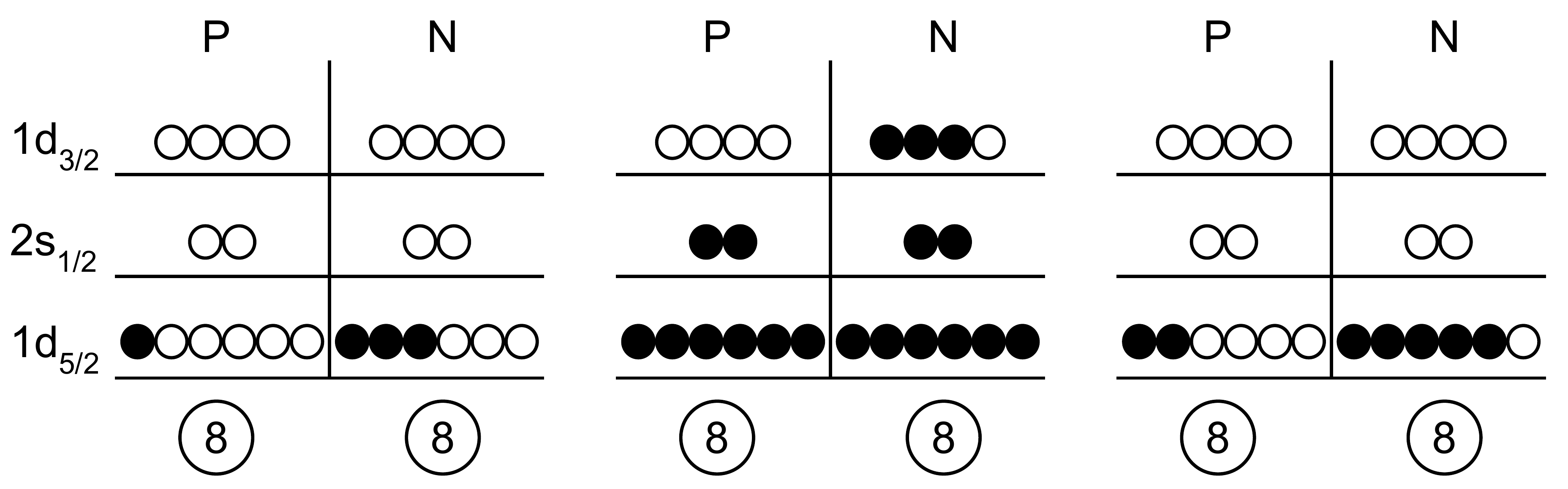}
\caption{Simple shell model diagrams illustrating degrees of freedom.  Like all of the nuclei included in our calculations, each of the three nuclei shown here has 8 protons and 8 neutrons in the core that are taken as inert, and no higher-energy single particle states beyond $1d_{3/2}$ are included.  Filled circles are valence particles, and empty circles are holes.  The left diagram shows $^{20}$F; this nucleus is excluded from our calculations for having too few protons.  The middle shows $^{35}$S, which is excluded for having too few neutron holes.  The right diagram shows $^{23}$Ne; this nucleus meets our degrees-of-freedom requirements, but we do not include isospin-lowering reactions for it, because the $^{23}$F daughter nucleus would have too few protons.}
\label{fig:dof}
\end{figure}

In order to ensure that our results are as realistic as possible, we set restrictions on which nuclei to compute.  Nuclear mass is a key factor in charged current weak processes; the parent and daughter masses are terms in $q_{if}$ in equations \ref{eq:phase_cap} and \ref{eq:phase_dec}.  Thus, we only consider nuclei that have experimentally measured mass.  Furthermore, a highly restricted model space (one which allows few single particle configurations) will not necessarily yield reliable results as there may simply not be enough basis states to accurately model the nucleus.  Therefore, we require that both the parent \emph{and} daughter nuclei have at least two degrees of freedom for \emph{each} of protons and neutrons in the $sd$ shell; the number of degrees of freedom is the lesser of valence particles and valence holes.  The core contains 8 protons and 8 neutrons, and the $sd$ shell can hold 12 nucleons of each species.  Therefore, the two-degrees-of-freedom requirement implies that both the parent and daughter nuclei must have at least 10 and at most 18 protons, and at least 10 and at most 18 neutrons; figure \ref{fig:dof} illustrates this.

To compute the energy levels and wave functions of the nuclei that meet our selection requirements, we use the USDB Hamiltonian \citep{br:2006}, as it has been demonstrated to reliably reproduce experimental nuclear energy levels and transition strengths \citep{rmb:2008}.  Once we have the wave functions, we compute the Gamow-Teller and Fermi matrix elements, defined by the following operators.

\begin{equation}
\label{eq:operators}
\begin{aligned}
\widehat{GT_\pm}&=\sum\limits_n \overrightarrow{\sigma}_n\widehat{\tau_\pm}_{n} \\
\widehat{GT_3}&=\sum\limits_n \overrightarrow{\sigma}_n\widehat{\tau_3}_{n} \\
\widehat{F_\pm}&=\sum\limits_n \widehat{\tau_\pm}_{n}
\end{aligned}
\end{equation}

The sums in equation \ref{eq:operators} are over nucleons, $\overrightarrow{\sigma}_n$ is the spin operator and $\overrightarrow{\tau}_n$ is the isospin operator on nucleon $n$.  Isospin raising, lowering, and component operators are all defined analogously with spin.  Again, in this work, we define protons as having isospin $+\frac{1}{2}$ and neutrons as having isospin $-\frac{1}{2}$, with the third component of isospin therefore given by $(Z-N)/2$.

The square of the matrix element for each operator between initial nuclear state $\vert i\rangle$ and final nuclear state $\vert f\rangle$ defines the strength of the corresponding transition.  We require the shell model code to compute the GT matrix elements, but the Fermi strength can be computed ``by hand'' by assuming that it is concentrated in the isobaric analog state of the daughter nucleus, or equivalently, that isospin is a good symmetry.  The Fermi strength is then simply

\begin{equation}
B_{if}^{F_\pm}=\vert\langle f\vert \widehat{T_\pm}\vert i\rangle\vert^2=T_i(T_i+1)-T_{3i}(T_{3i}\pm 1),
\end{equation}
where $\widehat{T}=\sum\limits_n \widehat{\tau_n}$ is the total nuclear isospin operator with eigenvalue $T$.  The total isospin raising/lowering operator $T_\pm$ has the selection rules that the total isospin must remain unchanged $(T_i=T_f)$, and the third component of isospin must change by $\pm 1$.  This implies, of course, that we neglect any effects of isospin symmetry breaking, whose influence on nuclear astrophysics has not been much explored. However, study in nuclear structure suggests that the influence may not be small (\citeauthor{wilkinson:1971} \citeyear{wilkinson:1971}; Kaneko et al 2017, private communication); this will be the topic of a future investigation.

Once the nuclear energy levels and transition strengths are computed, we replace any of our theoretical values with experimentally measured energies and $\beta$-decay (electron capture, etc.) strengths whenever the measured state can be matched to the corresponding theoretical state.  If a nucleus does not have experimental data for a particular transition but the mirror system does (same nuclear mass $A$, but proton and neutron numbers switched), we use the data from the mirror.  We use the following method to match states.

\begin{enumerate}
\item Match states in order of increasing experimental energy, beginning at the ground state.

\item If a measured state has exactly one theoretical state within $\sim 0.5$ MeV with the same spin and parity, it is a confident match.  If the theoretical energy is greater than 0.5 MeV away, it may be an unconfident match.

\item If there are multiple nearby theoretical and experimental states in equal quantity with the same spin and parity, confidently match them in ascending order.

\item In cases where there are more nearby theoretical states than experimental, preferentially match the lower-energy theoretical state unless the higher-energy state is much closer in energy.  This is an unconfident match.

\item If there are more nearby experimental states than theoretical with the same spin and parity, unconfidently match with the lower-energy experimental state.

\item If a measured state has uncertain spin/parity, unconfidently match with an appropriate theoretical state.  Inspecting measured gamma decays can help decide whether a theoretical state has the appropriate spin and parity.

\item When confidently matchable states become sparse, cease matching and, with the following exception, use only theoretical values above that energy.

\item Exception: States with isospin $T>T_3$ are matched independently of the rest of the states.  Since these can be the isobaric analog daughter states for super allowed transitions, they are both readily identifiable in experiments (so they may be confidently matched) and major contributors to weak rates.  They may have substantially higher energy than the other confidently matched states.
\end{enumerate}

Once states have been matched, we replace the theoretical energies with their experimental values.  We also replace the theoretical transition strengths with available measured strengths for all matched states.  If measured strength is unavailable or if a measured strength is available but either the parent or daughter state is unmatched, we rely on the theoretical results.

The GT$_\pm$ strengths computed from the usdb Hamiltonian tend to overestimate the experimental values by $\sim 40\%$, so we apply a uniform quenching factor of $0.764^2\approx 0.584$ to all of our theoretical GT$_{\pm,3}$ strengths \citep{rmb:2008}.  This is $\sim 1\%$ lower than the value of 0.59 used by \cite{oda-etal:1994}, but the differences between the Hamiltonian, techniques, and available experimental data of that work and here yield larger discrepancies.  Therefore, we are not concerned about this small disparity.  Fermi strengths are functions only of total nuclear isospin, so they are unquenched.

\subsection{Spectra}
\label{sec:spectra}

Equation \ref{eq:rate_generic} produces nuclear reaction rates, but we are interested here in neutrino energy spectra.  To obtain the spectra, we turn to the phase space factors, equations \ref{eq:phase_cap}, \ref{eq:phase_dec}, and \ref{eq:phase_deex}.  These integrals can be rewritten in terms of the outgoing neutrino energy.  The kernel of each integral--once multiplied by the physical constants and strength--can then be interpreted as the nuclear transition rate as a function of neutrino energy, and we are simply integrating over neutrino energy to get the total rate.  In other words, when written in terms of the neutrino energy, the kernels of the phase space integrals provide the energy spectrum of neutrinos originating from transitions from nuclear state $\vert i\rangle$ to nuclear state $\vert f\rangle$.  In charged lepton capture processes, the (anti-)neutrino energy is $w_\nu=-q_{if}+w_{e^\mp}$, so we can make the substitution $w_{e^\mp}=q_{if}+w_\nu$ in the kernel of equation \ref{eq:phase_cap} to obtain the spectrum

\begin{equation}
\label{eq:spec_cap}
S_{if}^{ec,pc}(w_\nu)=\left(\lambda_0^{GT_\mp}B_{if}^{GT_\mp}+\lambda_0^{F_\mp}B_{if}^{F_\mp}\right)w_\nu^2(q_{if}+w_\nu)^2G(Z,q_{if}+w_\nu)f_{e^\mp}(q_{if}+w_\nu)[1-f_{\nu,\overline{\nu}}(w_\nu)],
\end{equation}
where upper signs are for electron capture (neutrinos) and lower signs for positron capture (antineutrinos).  The (anti-)neutrino energy in charged lepton decays is $w_\nu=-q_{if}-w_{e^\mp}\rightarrow w_{e^\mp}=-q_{if}-w_\nu$, giving the spectrum

\begin{equation}
\label{eq:spec_dec}
S_{if}^{ed,pd}(w_\nu)=\left(\lambda_0^{GT_\pm}B_{if}^{GT_\pm}+\lambda_0^{F_\pm}B_{if}^{F_\pm}\right)w_\nu^2(-q_{if}-w_\nu)^2G(Z,-q_{if}-w_\nu)[1-f_{e^\pm}(-q_{if}-w_\nu)][1-f_{\overline{\nu},\nu}(w_\nu)],
\end{equation}
where upper signs are for electron emission (antineutrinos) and lower for positron emission (neutrinos).  Recall that the energies for charged lepton capture and emission are in units of electron mass and need to be converted to our units of choice.  The neutral current de-excitation spectrum is, up to the blocking factors, the same for neutrinos and antineutrinos.

\begin{equation}
\label{eq:spec_deex}
S_{if}^{de}(E_\nu)=\lambda_0^{GT_3}B_{if}^{GT_3}E_\nu^2(-Q_{if}-E_\nu)^2[1-f_\nu(E_\nu)][1-f_{\overline{\nu}}(-Q_{if}-E_\nu)]
\end{equation}

With our choices of units, the dimensions of the spectra will be neutrinos s$^{-1}$ MeV$^{-1}$.  However, because of our definition of $D$ in equation \ref{eq:cc_constants}, equations \ref{eq:spec_cap} and \ref{eq:spec_dec} are in neutrinos s$^{-1}$ m$_e^{-1}$, so we must divide by the electron mass in MeV in those equations to fix the units.  In our calculations, we also divide by the mass number $A$, so our tabulated spectra will ultimately have the units neutrinos s$^{-1}$ MeV$^{-1}$ baryon$^{-1}$.

Naturally, the neutrino spectrum from a nucleus in a particular initial state will be the sum of the spectra from transitions to all final states.  Moreover, the occupation probability of an initial state $\vert i\rangle$ is equal to the product of its spin degeneracy $2J_i+1$ and Boltzmann factor $e^{-E_i/T}$ divided by the nuclear partition function at temperature $T$.  The total tabulated neutrino spectrum per baryon due to reaction $X$ in a nucleus with mass number $A$ at temperature $T$ is then

\begin{equation}
S^X(E_\nu)=\frac{1}{A}\frac{1}{G(T)}\sum\limits_{i,f} \left( 2J_i+1\right)e^{-E_i/T}S_{if}^X(E_\nu),
\end{equation}
where $G(T)$ is the nuclear partition function, $E_i$ is the initial excitation energy of the parent nucleus, and the sum is over initial and final nuclear states.  In these calculations, we take the charged leptons to be in Fermi-Dirac distributions, which are functions of the temperature, density, and electron-to-baryon ratio $Y_e$.  We assume that neutrinos stream freely away, implying that neutrino blocking is zero.

We computed the charged current neutrino spectra from 0 to 30 MeV at 0.5 MeV resolution and the neutral current deexcitation neutrino spectra from 0 to 20 MeV at 0.1 MeV resolution on the FFN temperature-density grid \citep{ffn:1980,ffn:1982a,ffn:1982b,ffn:1985}.  Neutrinos from charged lepton captures with $Q_{if}<0$ are nearly monoenergetic at low temperature and density, as the leptons will have little kinetic energy.  So for each nucleus, we did the following.  If the ground state of the nucleus has allowed charged lepton captures with $Q_{if}<0$, we computed the spectral density just above $E_\nu=-Q_{if}+m_e$.  The Coulomb correction favors low energy electrons, so for electron captures, we computed the spectrum at $E_\nu=-Q_{if}+m_e+10^{-5}$ MeV.  By contrast, positrons are repelled by the nucleus, and the Coulomb correction therefore tends to spread positron capture neutrinos up in energy; for positron captures, we computed the spectrum at $E_\nu=-Q_{if}+m_e+0.01$ and $+0.1$ MeV.  Because this is such a precise correction, we only explicitly include captures between experimentally matched states.  In a few rare cases, we include capture from a low-lying first excited parent state to a daughter ground state.  At higher temperatures and densities, the charged lepton distributions spread to energies greater than our spectrum resolution, with a concomitant spreading of the capture peaks.  These peaks are discussed further with illustrative figures in section \ref{sec:discussion}.

\section{Discussion and Usage Notes}
\label{sec:discussion}

Our 0.5 MeV neutrino energy resolution effectively captures spectral features with width greater than the resolution, but two problems arise when sharp features occur between the tabulated neutrino energies.  First, sharp peaks in the spectrum may be completely excluded and hence missed by users of the spectra.  Second, the total neutrino production and energy loss rates obtained by integrating the spectra may not always be reliable.  Here, we present some considerations and workarounds for these shortcomings.

From equation \ref{eq:spec_deex}, neutral current deexcitation neutrino spectra take a polynomial form for each transition, so they tend to have broad peaks.  This fact, coupled with the comparatively fine energy resolution of our computations, means that these spectra can be reliably integrated numerically with a simple rectangular or trapezoidal method, and these spectra are generally not subject to the failure modes described above.

Three factors limit the accuracy of simple numerical integration of the charged current spectra.  First, at temperature-density points with temperature below $\sim 0.1$ MeV and density $\rho Y_e$ below $\sim 10^6$ g/cm$^3$, peaks from captures with $Q_{if}<0$ are \emph{much} narrower than the neutrino energy resolution.  Second, even in the absence of sharp capture peaks, the resolution is somewhat low, so a simple integration method may produce significant errors.  Third, for reactions that take a parent nucleus to a daughter with a higher mass, the neutrino peak may be at lower energy than the resolution.  For these reasons, we recommend deferring to the results of \cite{oda-etal:1994} for total neutrino rates and energy losses.

The nuclei in the following examples were chosen simply because they have comparatively few basis configurations in the $sd$ shell, which enables fast computation of the enhanced-precision spectra used for illustration.  They are not necessarily singularly interesting, though they will be present in the nuclear statistical equilibrium environment in the cores of highly evolved massive stars.  In the figures below, crosses show explicitly the neutrino energies for which spectral density was computed.

\begin{figure}
\centering
\includegraphics[scale=0.4]{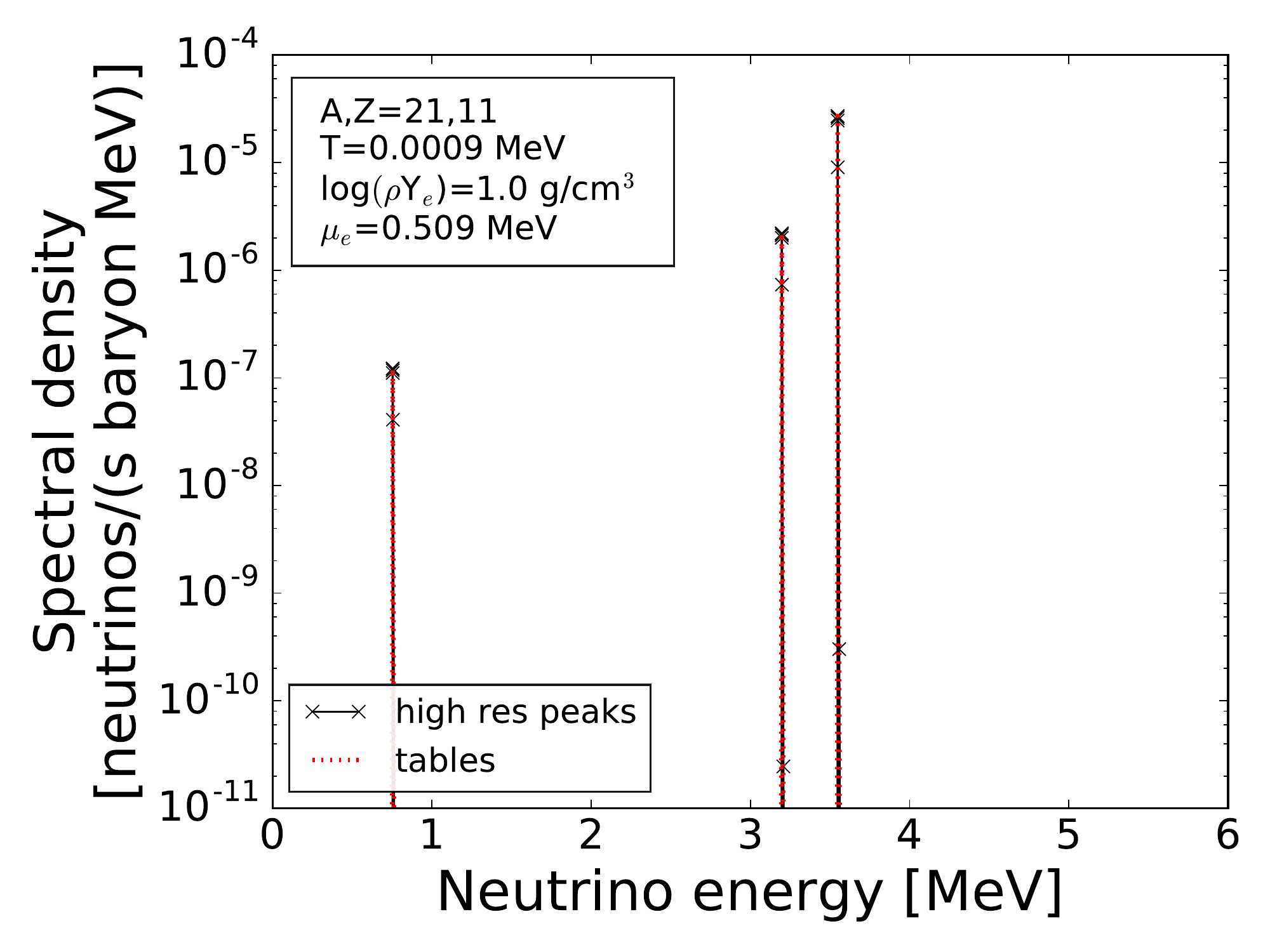}
\includegraphics[scale=0.4]{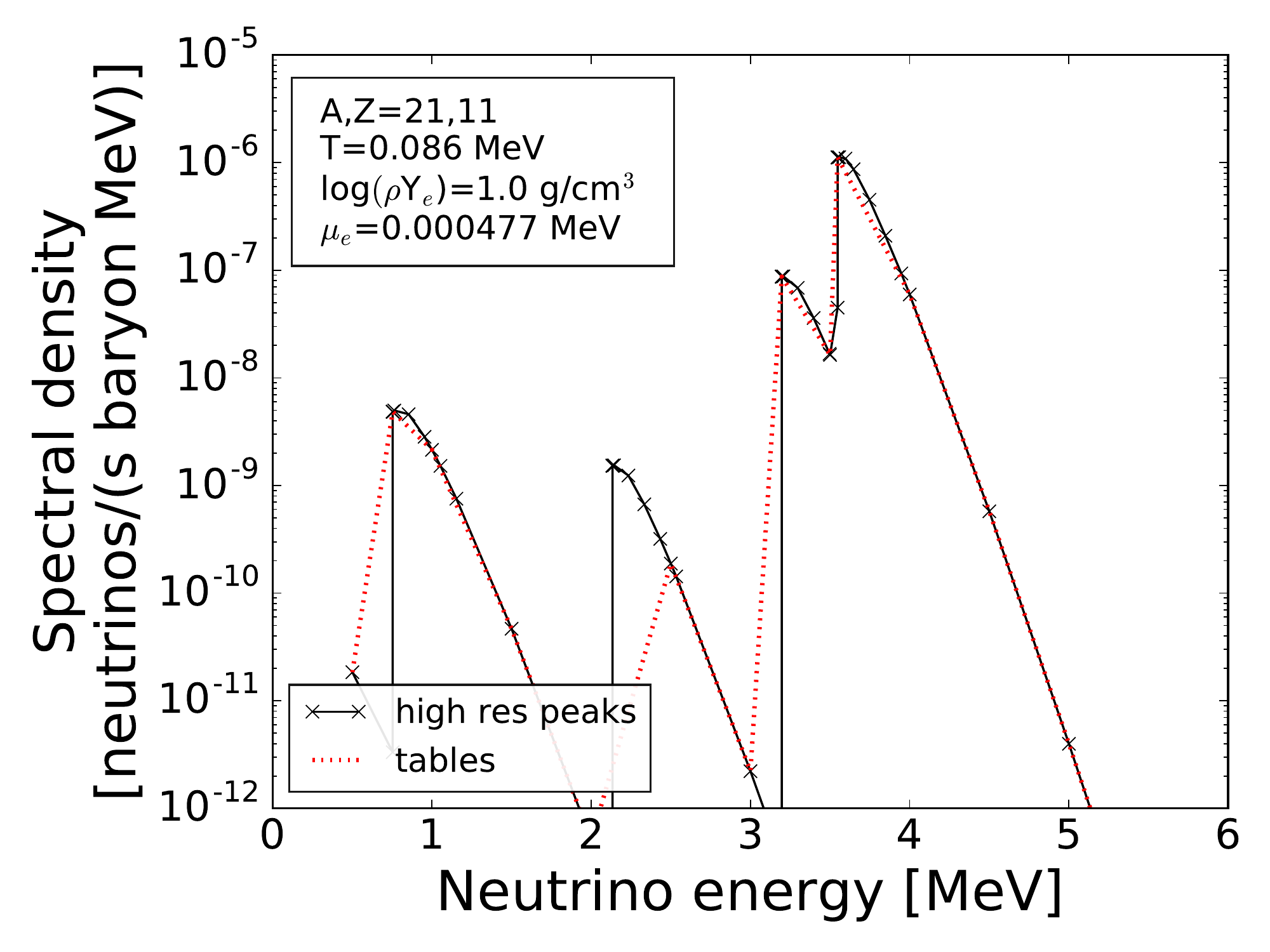}\\
\includegraphics[scale=0.4]{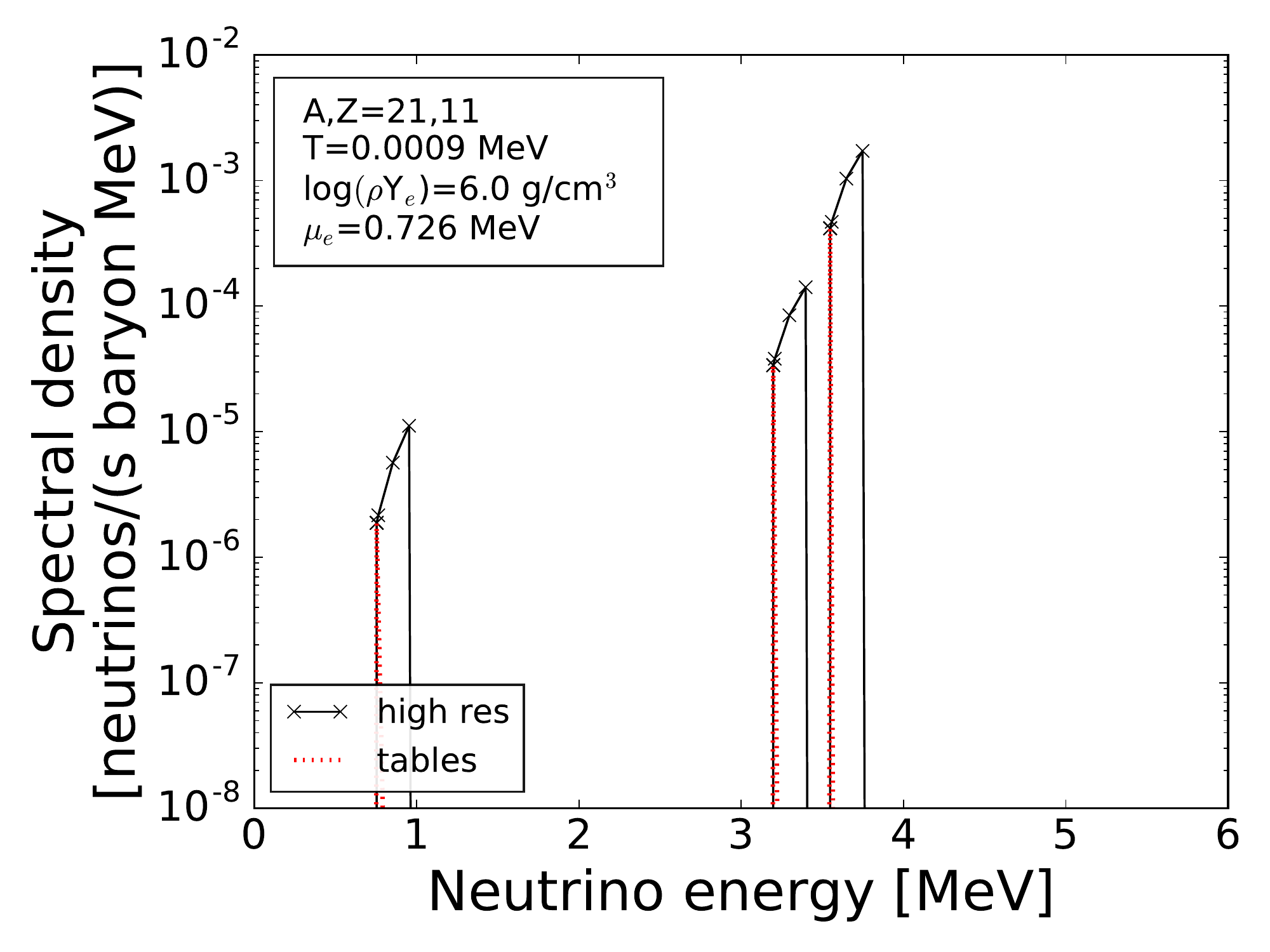}
\includegraphics[scale=0.4]{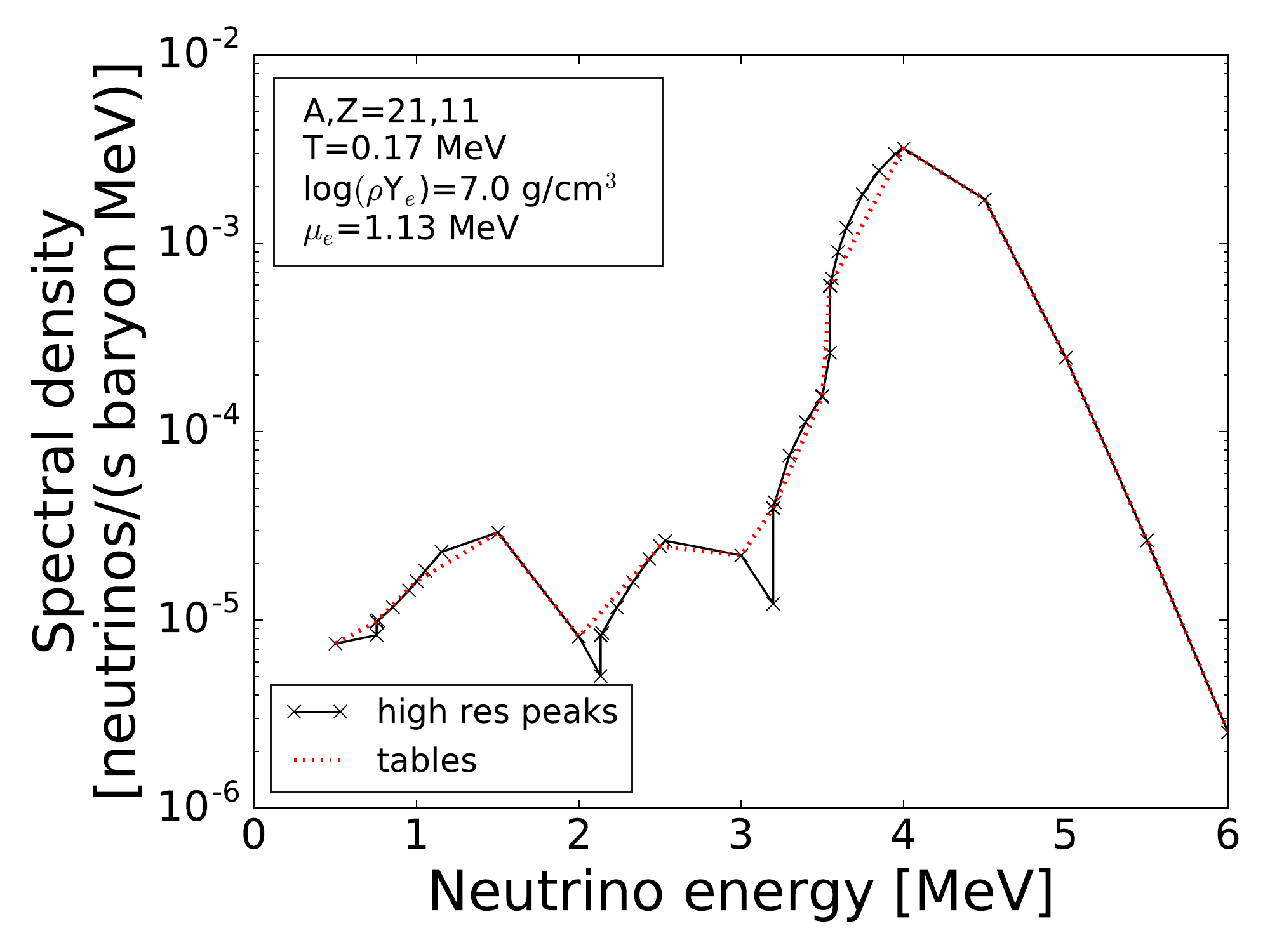}
\caption{$^{21}$Na electron capture spectra.  The crosses show where the spectral density was computed for the spectra with high resolution around ground state electron capture peaks, with solid black lines connecting (``high res peaks'').  The dotted red lines connect the points that appear in our spectrum data tables (``tables'').  In order from left to right, top to bottom, the panels show extremely low temperature and density, modest temperature with extremely low density, extremely low temperature with modest density, and realistic temperature and density for a stellar core during silicon burning.  In each panel, $\mu_e$ is the electron chemical potential.}
\label{fig:e_cap_spect}
\end{figure}

\begin{figure}
\centering
\includegraphics[scale=0.4]{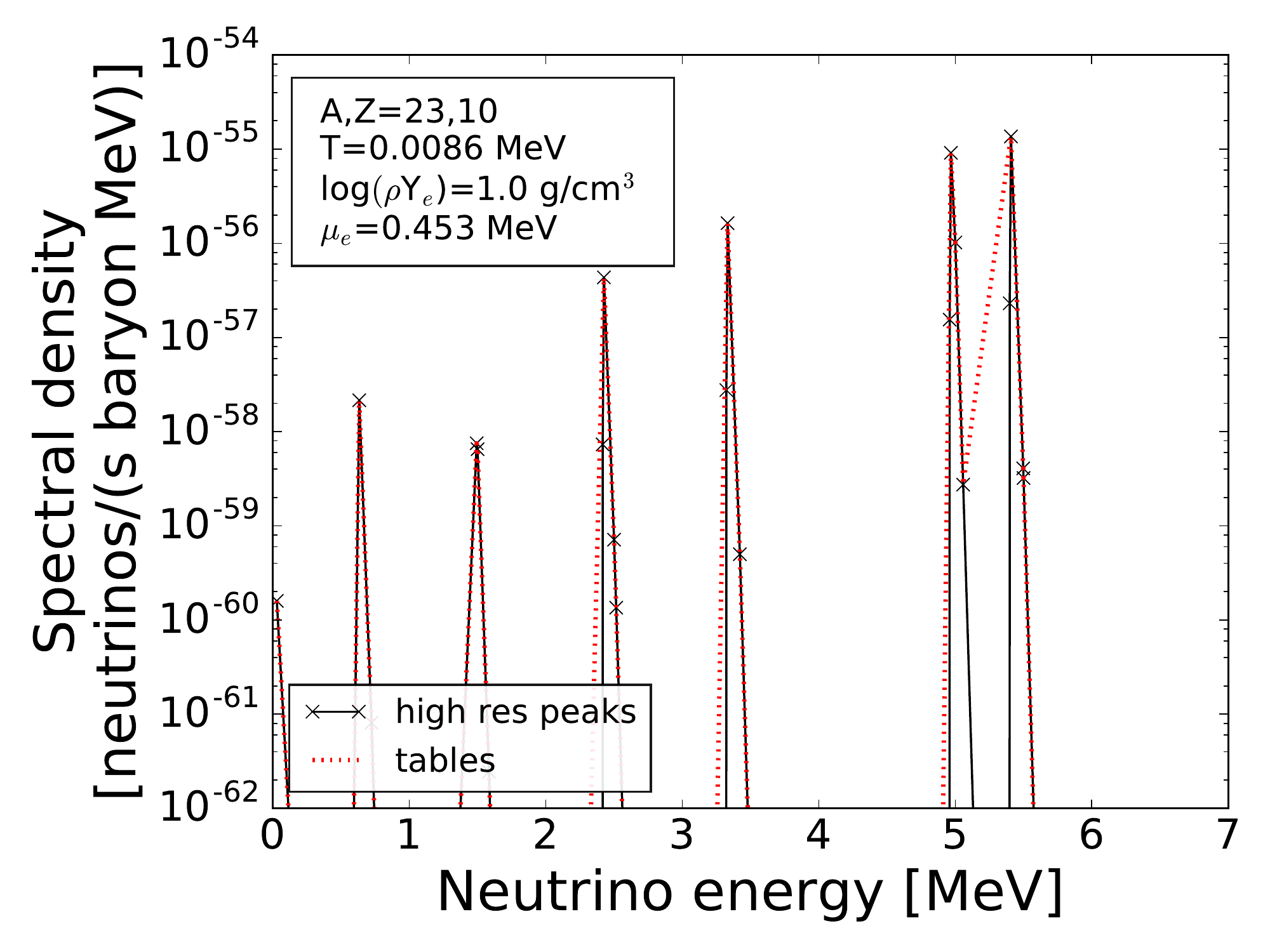}
\includegraphics[scale=0.4]{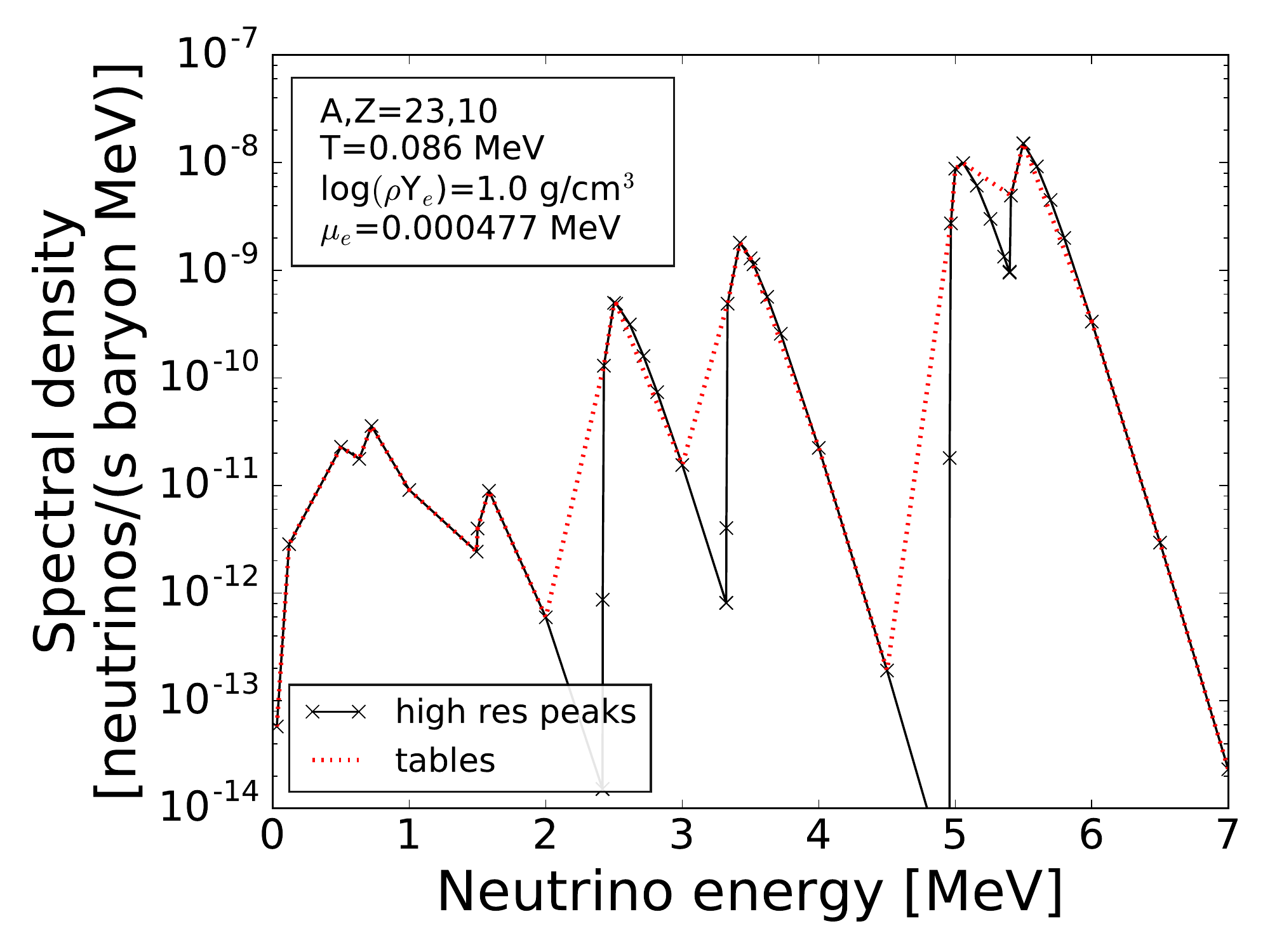}\\
\includegraphics[scale=0.4]{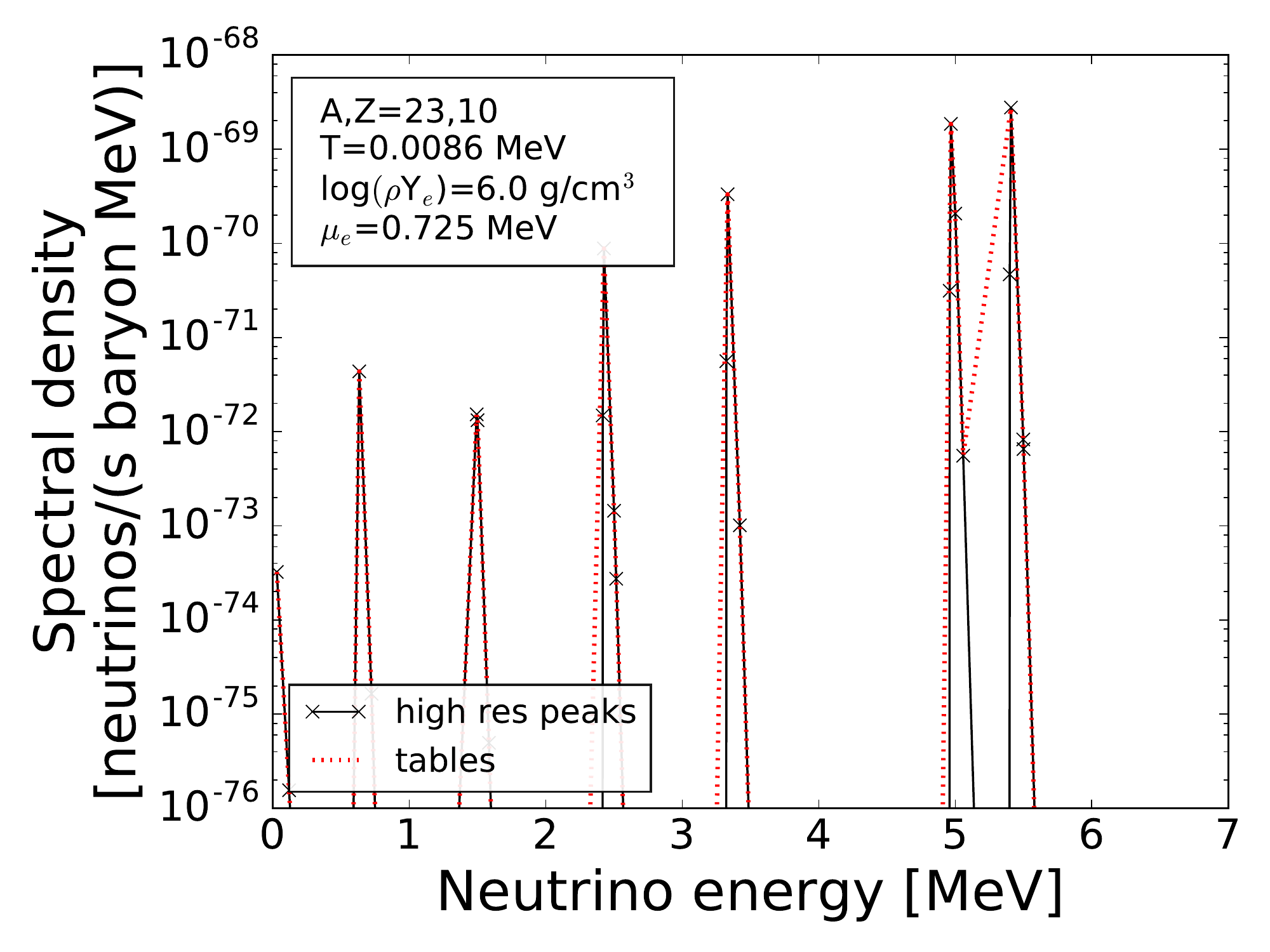}
\includegraphics[scale=0.4]{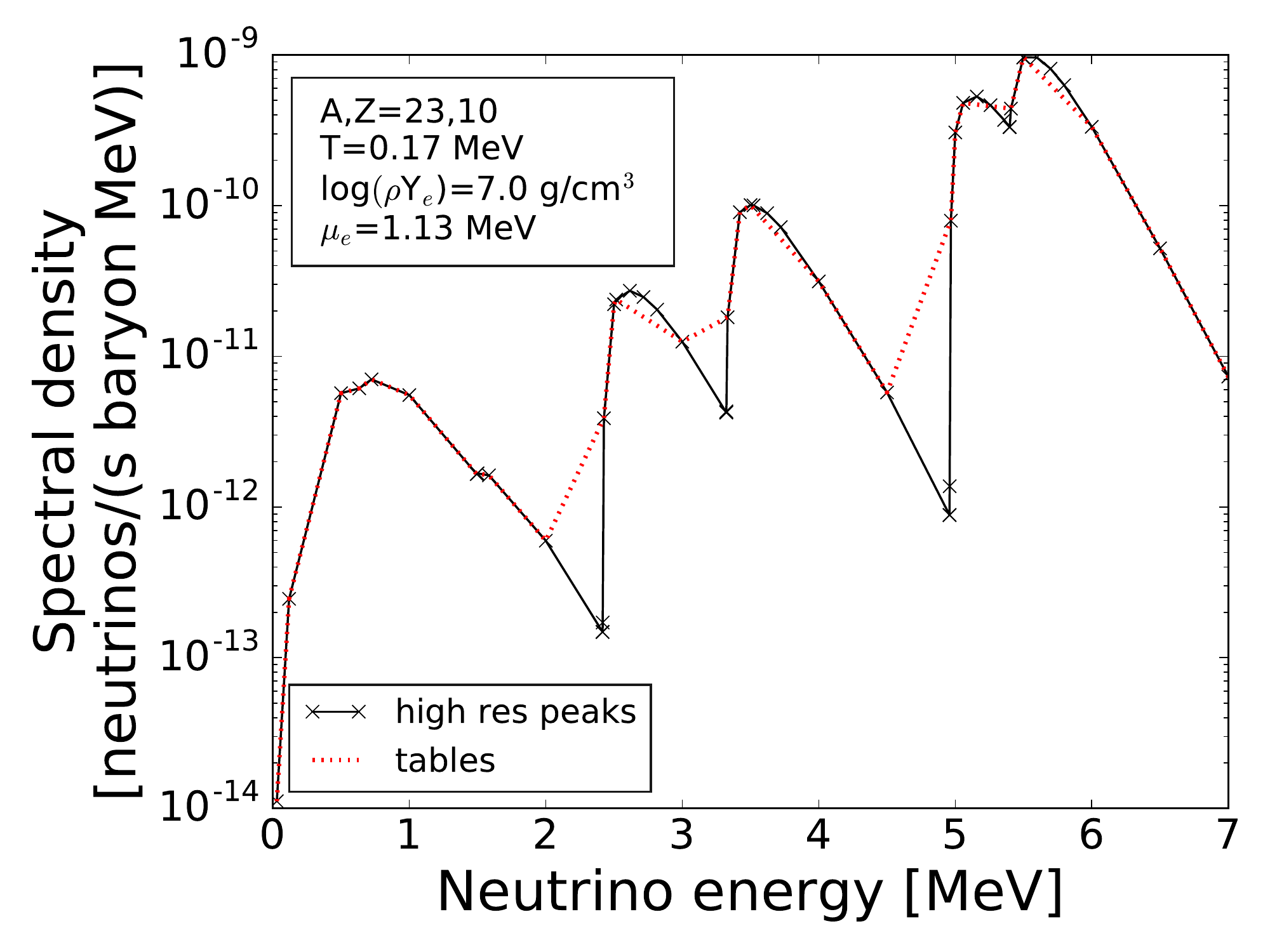}
\caption{$^{23}$Ne positron capture spectra.  The lines and panels are as in figure \ref{fig:e_cap_spect}, though here the low temperature panels are 10 times higher, because at lower temperatures there is no significant population of positrons.}
\label{fig:p_cap_spect}
\end{figure}

Figures \ref{fig:e_cap_spect} and \ref{fig:p_cap_spect} illustrate peaks in the charged lepton capture spectra.  The figures show the spectra included in our tables and spectra with enhanced precision around the capture peaks; the enhanced precision peaks will not be tabulated and are only included here for illustrative purposes.  Figure \ref{fig:e_cap_spect} shows neutrino spectra from electron capture on $^{21}$Na at a variety of temperatures and densities, and figure \ref{fig:p_cap_spect} shows positron capture on $^{23}$Ne.  The upper left panel of each figure shows extremely low temperature and density; in these conditions, the peaks are extremely sharp, so increased precision provides little or no benefit.  The upper right panels show modest temperatures and extremely low density; at this temperature, the peaks are broad enough that they are adequately traced by the low resolution spectra.  The lower left panels show extremely low temperature and modest density; here, the positron capture peaks are unchanged in shape, but the degenerate electron Fermi sea has caused the electron capture peaks to broaden and move up in energy.  This effect is \emph{not} reflected in our tables.  The lower right panels show a realistic temperature and density in a silicon-burning stellar core; again, the low resolution spectra adequately trace the peaks.

An interesting feature appears in figure \ref{fig:e_cap_spect} at 2.0-2.5 MeV as the temperature increases.  This peak is due to electron capture from the 0.332 MeV first excited state of $^{21}$Na to the 1.746 MeV second excited state of $^{21}$Ne.  Because this peak is not due to capture on a ground state, it is \emph{not} explicitly included in our spectra.  The upper right panel of figure \ref{fig:e_cap_spect} demonstrates that there will be conditions where an excited state generates a significant peak that our tabulated spectra will not fully capture (though, as noted, there are a few exceptions where we include capture peaks from excited states).  However, the lower right panel shows that at moderate temperatures and densities, the peaks become broad enough to be reasonably represented.

\begin{figure}
\centering
\includegraphics[scale=0.4]{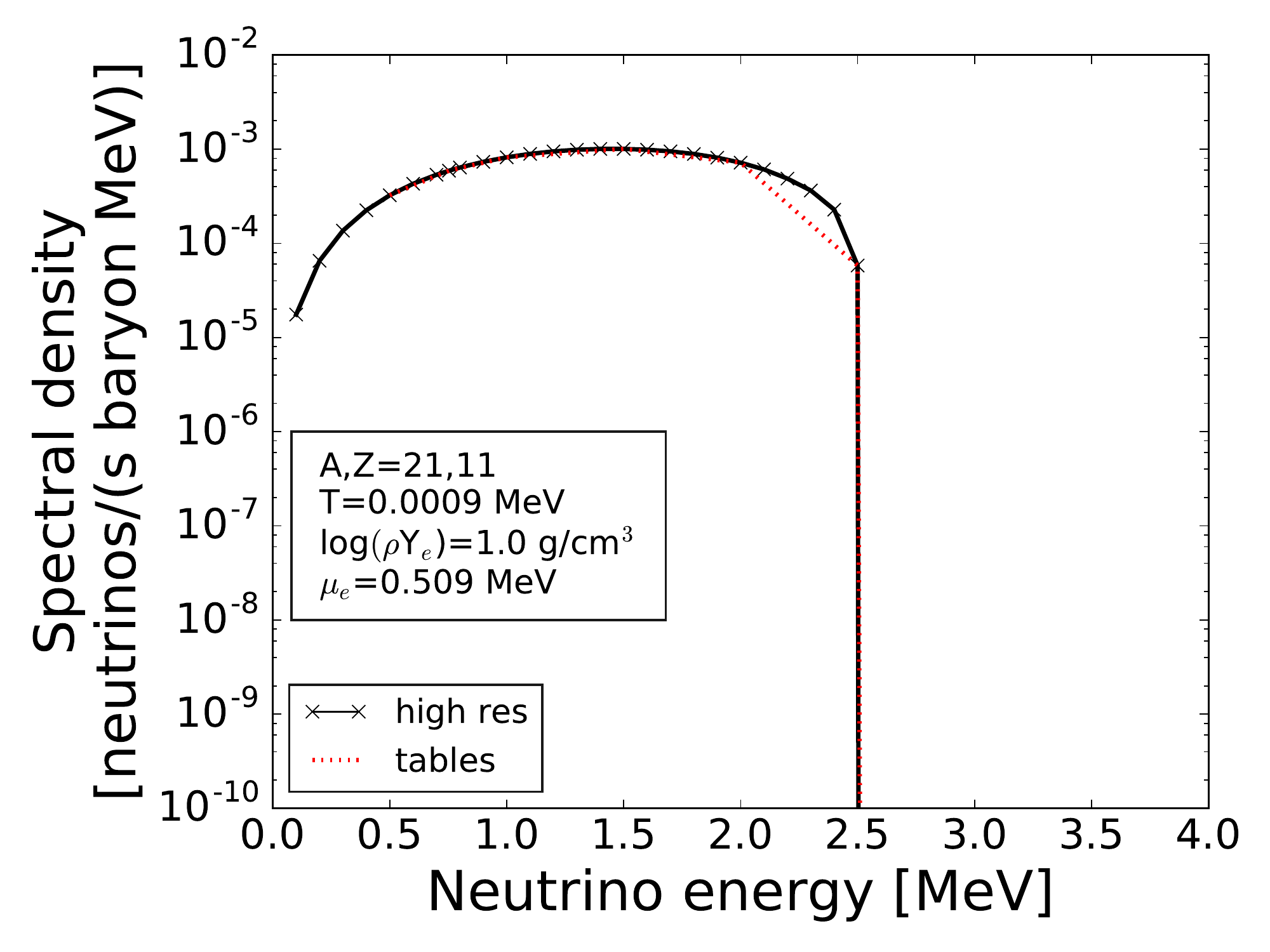}
\includegraphics[scale=0.4]{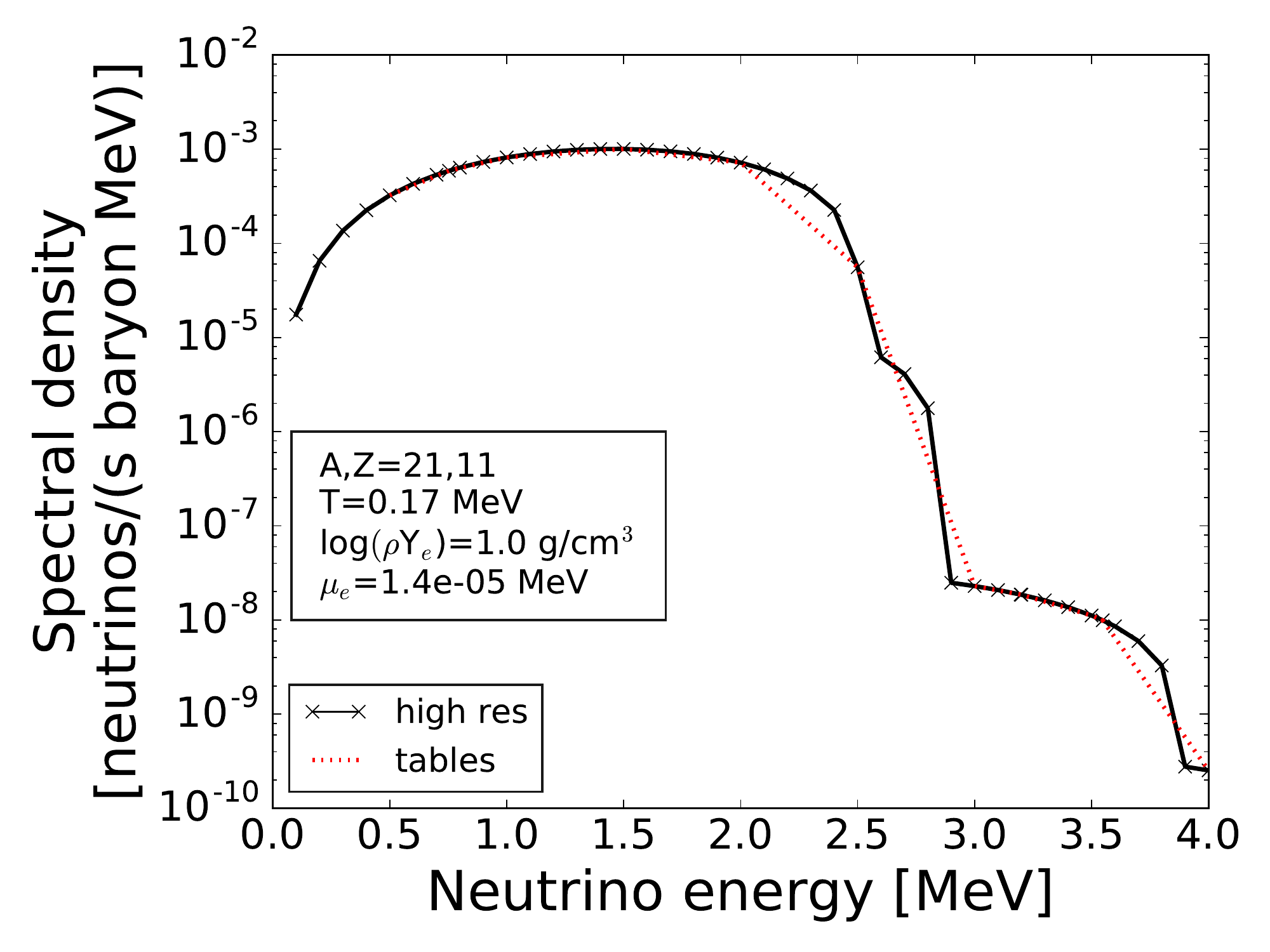}\\
\includegraphics[scale=0.4]{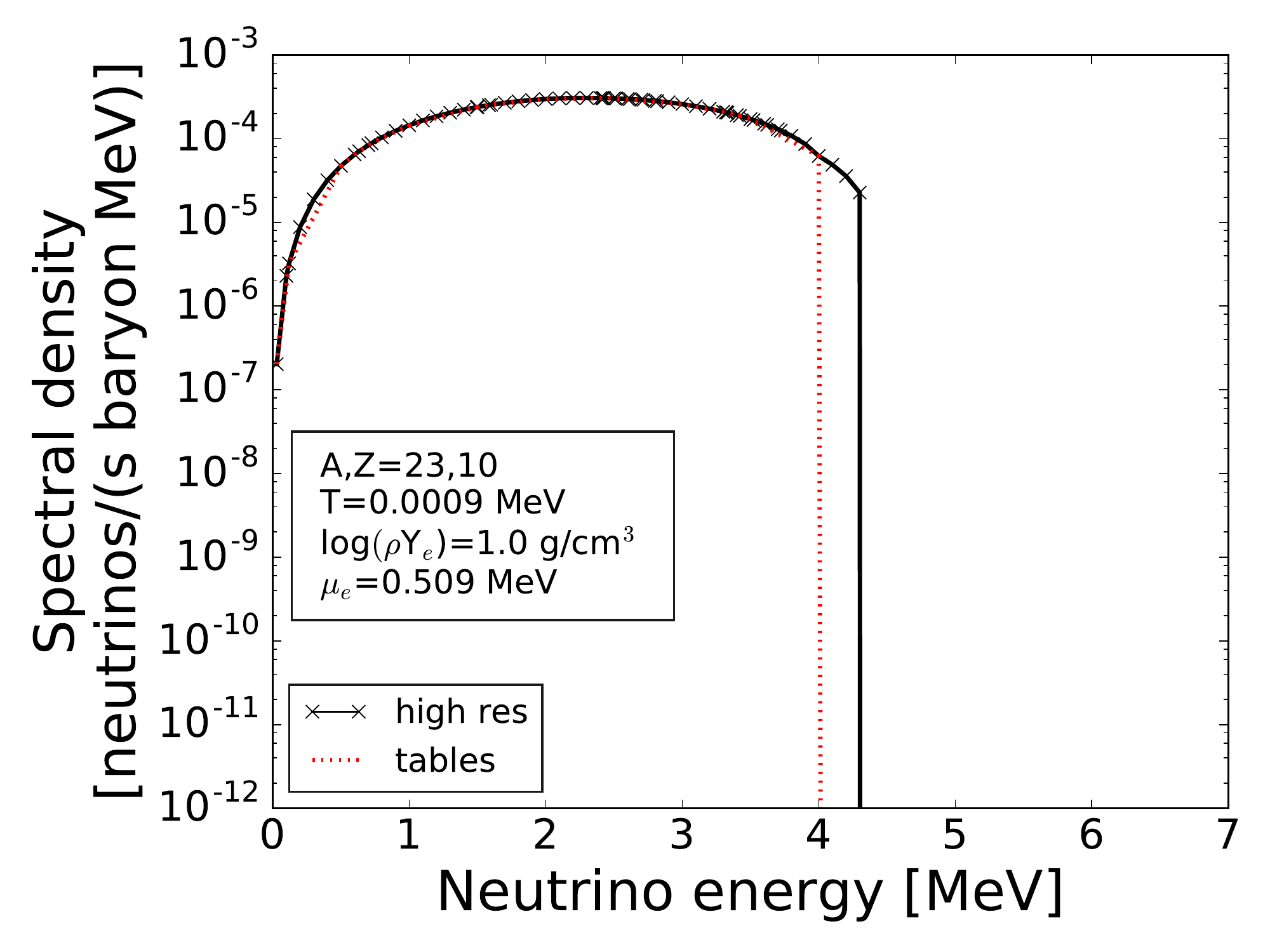}
\includegraphics[scale=0.4]{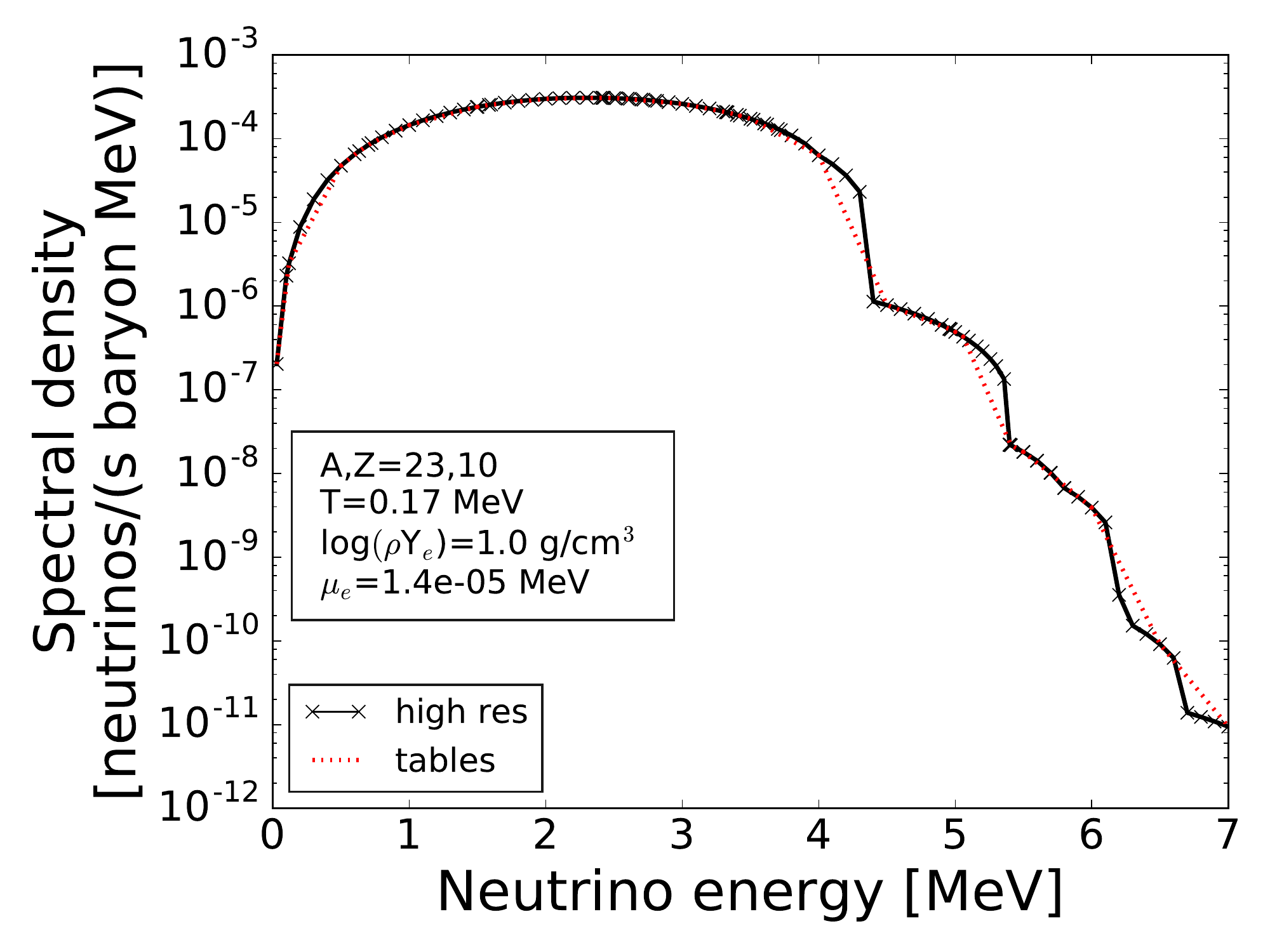}
\caption{$^{21}$Na positron emission (upper) and $^{23}$Ne electron emission (lower) spectra at extremely low density.  The black lines are at high resolution, with crosses showing explicitly where the spectral density was computed, and the red lines show the tabulated spectra.  The left panels show extremely low temperature and the right moderate temperature.}
\label{fig:dec_spect}
\end{figure}

Figure \ref{fig:dec_spect} shows positron emission neutrino spectra for $^{21}$Na and electron emission antineutrino spectra for $^{23}$Ne at low density and at low and moderate temperature.  In all cases, our tables trace the spectra quite well, with the only irregularities occurring near the drops at high energy; these drops occur at the maximum neutrino energy for specific parent-state-to-daughter-state transitions.

\begin{figure}
\centering
\includegraphics[scale=0.4]{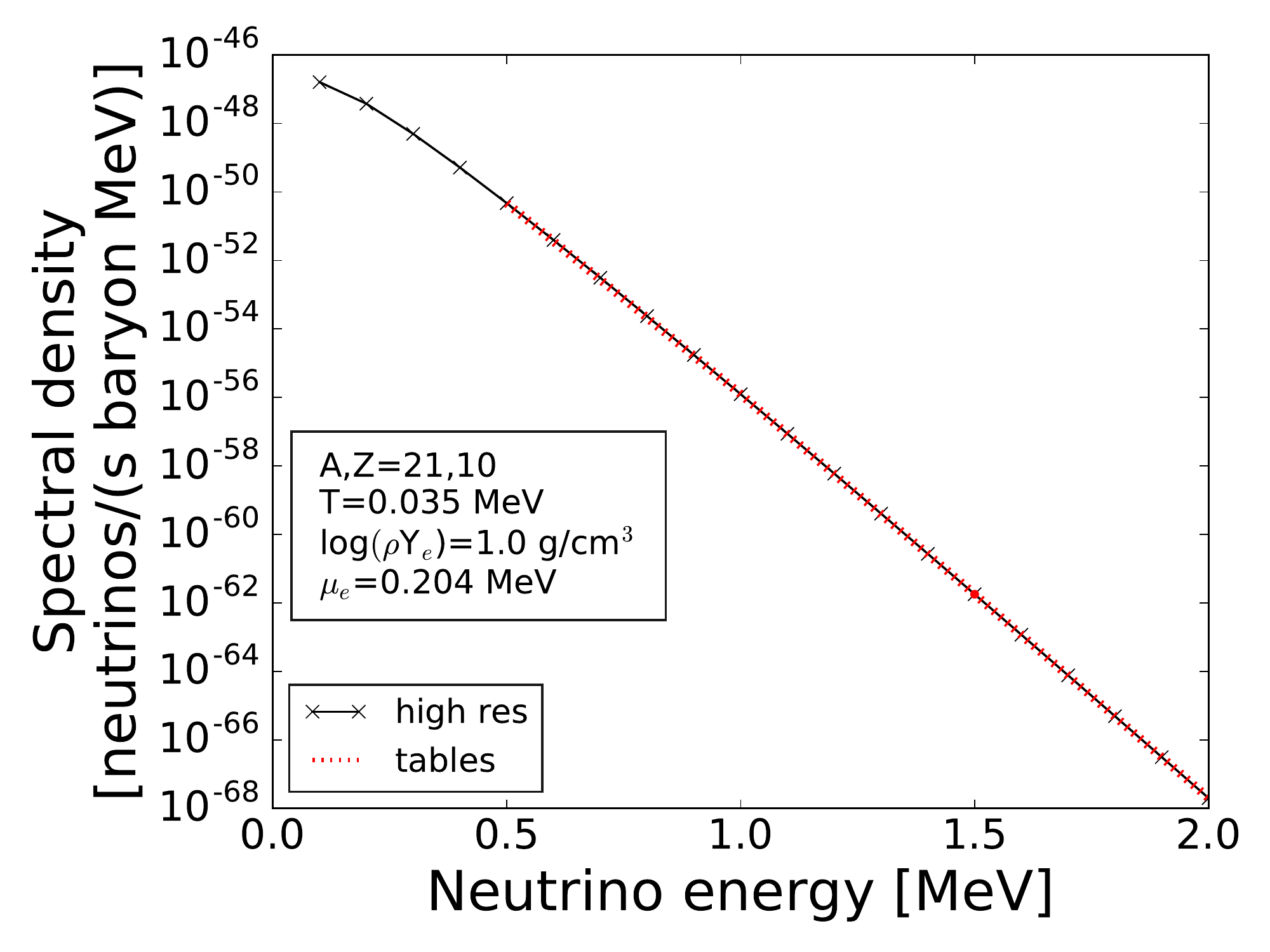}
\includegraphics[scale=0.4]{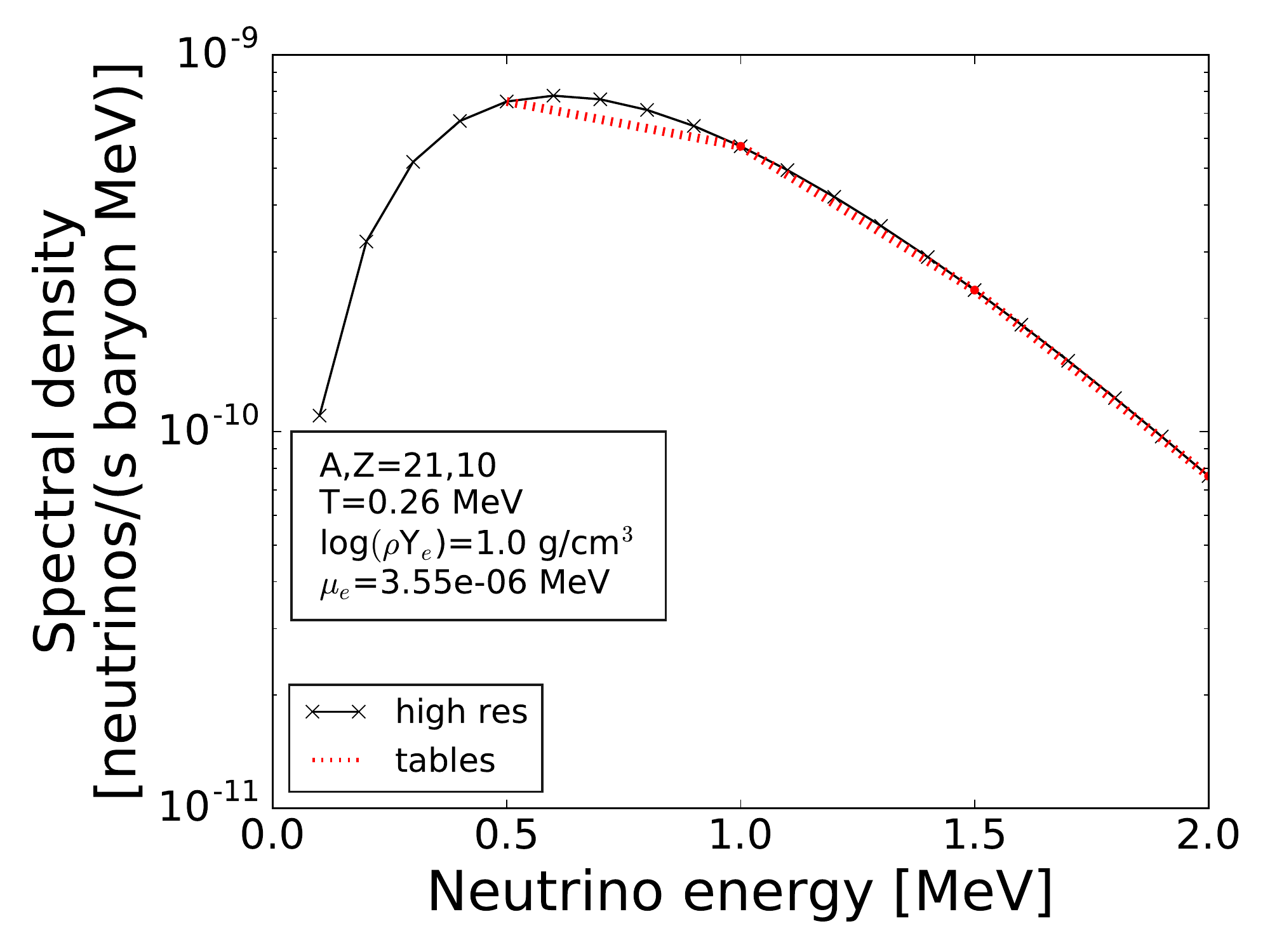}
\caption{$^{21}$Ne positron capture spectra.  The lines are as in figure \ref{fig:dec_spect}.  The left panel shows low temperature, and the right panel shows relatively high temperature.}
\label{fig:up_cap_spect}
\end{figure}

Figure \ref{fig:up_cap_spect} shows $^{21}$Ne positron capture spectra at low density and at low and high temperature.  In this case, the daughter $^{21}$Na nucleus has greater mass than parent.  At low temperature, the capture peak falls well below the resolution energy of our tables and is therefore not present at all.  The temperature and/or density must be fairly high before our tables will adequately represent such peaks.

In all of our capture spectra, sharp peaks present the greatest difficulty.  The high energy tails above the charged lepton chemical potential are smooth, so the 0.5 MeV resolution represents them adequately.  The pitfalls associated with peaks mean, however, that we must address the question of how many neutrinos are in each energy bin in the capture channels.  The simplest method is to scale the spectra by comparing with \cite{oda-etal:1994}: using your preferred numerical technique, integrate a spectrum to obtain the total reaction rate, then scale the spectrum by the ratio of the \cite{oda-etal:1994} rate to the computed rate.

\begin{equation}
^{\dagger}S^X(E_\nu)=\frac{\Lambda^{X}_{Oda}}{A\Lambda^{X}_{spec}}S^{X}(E_\nu)
\end{equation}

Here, $\Lambda^{X}_{Oda}$ is the \cite{oda-etal:1994} rate and $\Lambda^{X}_{spec}$ is the integrated rate computed from $S^X(E_\nu)$.  The factor of nuclear mass number $A$ accounts for the fact that our spectra are per baryon, while the \cite{oda-etal:1994} tables are per nucleus.  This method should produce satisfactory results for most spectra.  For sharply peaked spectra like those in the upper left of figures \ref{fig:e_cap_spect} and \ref{fig:p_cap_spect}, it is tantamount to assigning a width to the peaks.  Oddly-shaped peaks like those in the lower left of figure \ref{fig:e_cap_spect} can have tabulated values that are much lower than the actual peak values.  However, examining the spacing of the crosses on the lines demonstrates that the shapes of the capture peaks are determined principally by the charged lepton distributions (crosses indicate where the spectral density was computed, and for each peak it was computed at the same relative energy).  Thus, at a given temperature and density, the peaks have the same shape, and the spectrum can therefore be rescaled without affecting the \emph{relative} number of neutrinos in each peak.  For smooth spectra, rescaling simply serves as a correction factor for the integration.  Spectra that peak below the resolution cutoff, as in figure \ref{fig:up_cap_spect}, cannot be easily remedied by comparison with \cite{oda-etal:1994}, and the spectral densities must simply be taken as given.

To illustrate this method, consider the spectrum shown in the upper-left of figure \ref{fig:e_cap_spect}.  A trapezoidal integration of our tabulated $^{21}$Na spectra at $T\sim 0.001$ MeV and $\rho Y_e=1.0$ g/cm$^3$ yields an electron capture rate of $1.521\times 10^{-4}$ per nucleus per second, while \cite{oda-etal:1994} provides a rate of $3.048\times 10^{-7}$.  This indicates that the peaks in that spectrum are exceedingly narrow, and the total number of neutrinos produced per second per baryon in, for example, the peak near 3.55 MeV (the largest peak in that spectrum) is not $0.5 ~{\rm MeV} \times 2.6643\times 10^{-5} ~{\rm neutrinos/s ~baryon ~MeV}\approx 1.3\times 10^{-5} ~{\rm neutrinos/s ~baryon}$ (the trapezoidal result), but rather is smaller by a factor of $\frac{3.048\times 10^{-7}}{1.521\times 10^{-4}}\approx 2.0\times 10^{-3}$; we can effectively reproduce this by rescaling the entire spectrum by $2.0\times 10^{-3}$ and not further worrying about the effects of the resolution.  As a rule, because they are principally determined by the charged lepton distributions, capture spectra will consist either of sharp peaks or broad peaks--all with the same shape--with essentially no mixing of the two.  Consequently, we need not be concerned that a single large feature will dominate the correction factor and give erroneous results for other peaks in the spectrum.  This approach will yield reliable results for all of our tabulated spectra except where the peak occurs below the spectrum resolution energy (as indicated above and shown in figure \ref{fig:up_cap_spect}), and possibly when multiple extremely sharp capture peaks occur between the usual 0.5 MeV neutrino energy increments.  In the latter case, simply choose a single width which, when applied equally to every peak in the spectrum, produces the \cite{oda-etal:1994} result.



At modestly high temperatures and densities, where the spectra are relatively smooth, rates computed from integration of these spectra are expected to agree with \cite{oda-etal:1994}.  For example, trapezoidal integration of the $T=0.17$ MeV, $\rho Y_e=10^7$ g/cm$^3$ $^{21}$Na electron capture spectrum (lower-right of figure \ref{fig:e_cap_spect}) yields a rate of 0.0578 neutrinos/s nucleus, and \cite{oda-etal:1994} gives 0.0575.  However, due to differences in procedure, choice of Hamiltonian, available experimental data, etc, there may remain some disagreement.  For the sake of consistency, deferring to \cite{oda-etal:1994} and scaling the spectra so that numerical integration reproduces the \cite{oda-etal:1994} rates may be preferable in some applications.

Because our neutrino spectra are essentially transition rates to states with specific outgoing neutrino energies, it is easy to account for neutrino blocking of charged current interactions when conditions are such that there is a substantial population of neutrinos in the environment.  To adjust for neutrino blocking, simply scale the spectrum by the neutrino blocking factor as a function of neutrino energy.  Note that a spectrum must be adjusted to agree with \cite{oda-etal:1994} \emph{before} the blocking is applied.

\begin{equation}
\label{eq:spec_blocked}
S_{blocked}^X(E_{\nu,\overline{\nu}})=S^X(E_{\nu,\overline{\nu}})[1-f_{\nu,\overline{\nu}}(E_{\nu,\overline{\nu}})]
\end{equation}

This technique will \emph{not} precisely apply to blocking of neutral current neutrinos.  Imagine an environment where the neutrinos are completely blocked above 5 MeV, completely unblocked below 5 MeV, and the antineutrinos are completely unblocked everywhere (this is an exaggeration of a collapsing supernova core at the onset of neutrino trapping).  In this environment, the neutrino spectrum would be the same as the unblocked spectrum below 5 MeV and identically zero above.  However, the antineutrino spectrum does not remain unchanged, despite being unblocked, for the following reason.  If a neutrino pair is emitted with an energy of, say, 8 MeV, then the neutrino can have at most 5 MeV due to blocking.  However, this implies that the antineutrino must have \emph{at least} 3 MeV of energy.  Because neutrino pairs are emitted at many different energies, neutral current neutrino spectra cannot be easily modified to account for neutrino blocking.

\begin{figure}
\centering
\includegraphics[scale=0.4]{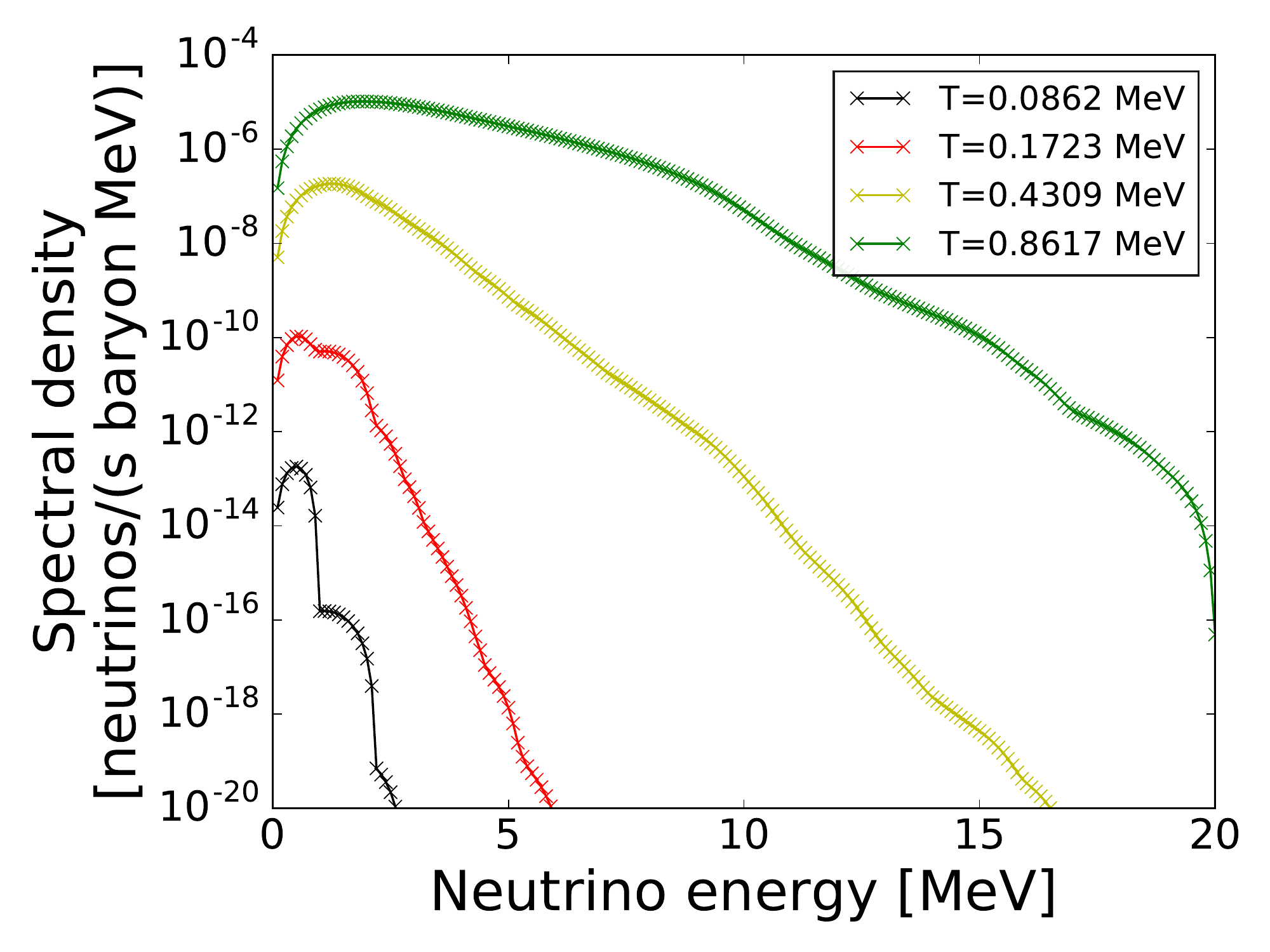}
\caption{$^{25}$Na neutral current deexcitation neutrino spectra.  The antineutrino spectra are identical, and all three flavors are equally present, so the spectrum for each flavor of each species will be one third of that shown here.}
\label{fig:nc_dec_spect}
\end{figure}

In the neutral current deexcitation channel, the 0.1 MeV resolution traces the neutrino spectrum closely enough to allow for reliable numerical integration.  Similar to charged current decays, individual transitions have broad spectra peaked at half the transition energy, so all of the important features are captured.  Figure \ref{fig:nc_dec_spect} shows neutral current deexcitation neutrino spectra for $^{25}$Na, given here as the sum of all three neutrino flavors.  The deexcitation spectra are \emph{not} the sum of neutrinos and antineutrinos; rather, they represent both the neutrino spectra and the antineutrino spectra, as the spectra of both species are identical.  Furthermore, because this reaction channel is blind to flavor, all three flavors of neutrino will be equally present.  Thus, the spectrum for each flavor of neutrino and antineutrino will be one third of the spectrum shown.  The sharp cutoff near 20 MeV is due to the omission of nuclear initial energy bins above 20 MeV.

\begin{figure}
\centering
\includegraphics[scale=0.4]{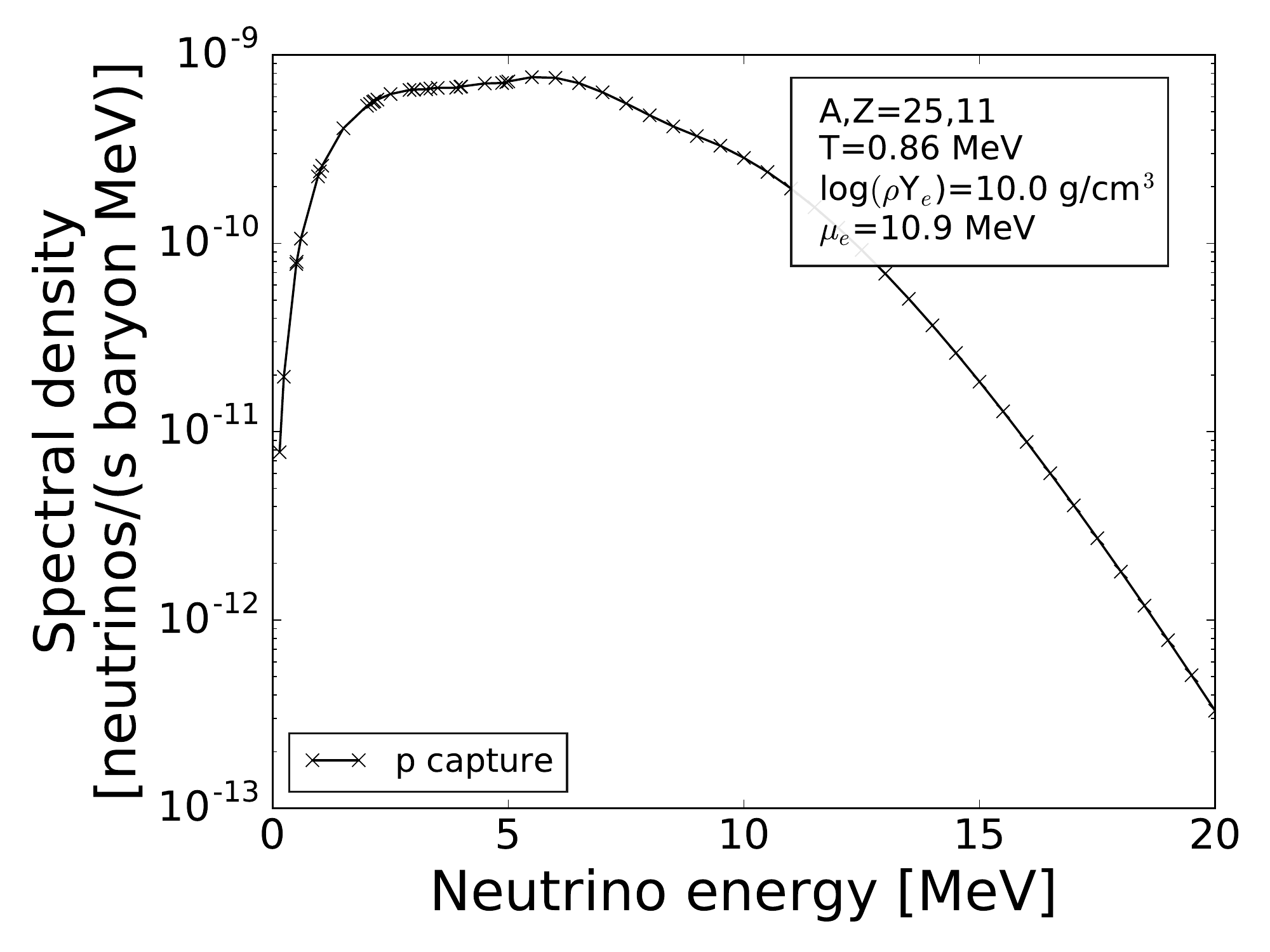}
\includegraphics[scale=0.4]{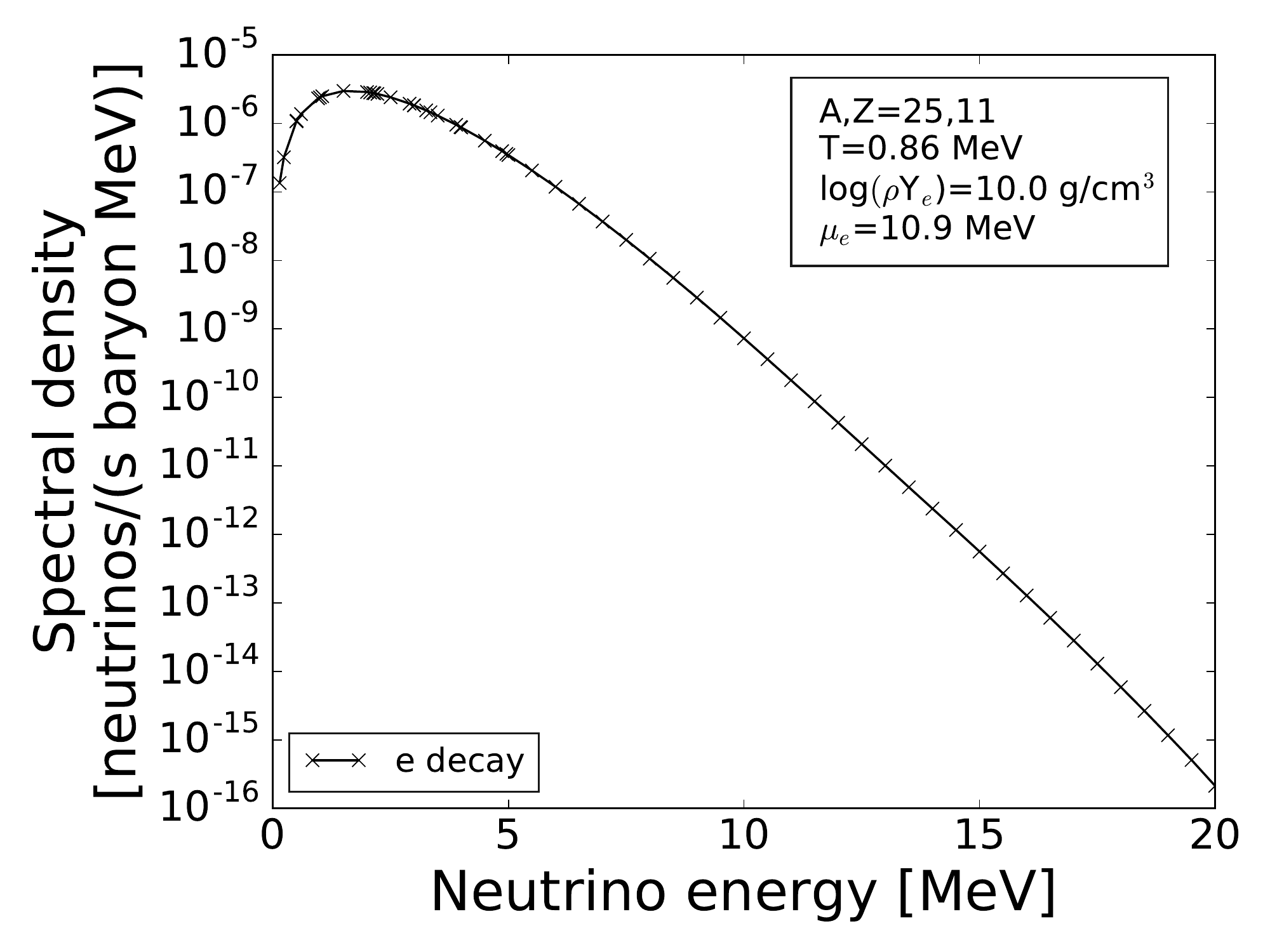}
\caption{$^{25}$Na isospin-raising reaction neutrino spectra at temperature and density typical of early core collapse.  The left panel shows positron capture, and the right panel panel electron emission.}
\label{fig:na_nubar_spect}
\end{figure}

Figure \ref{fig:na_nubar_spect} shows the antineutrino spectra from positron capture and electron emission in $^{25}$Na at temperature and density realistic for the early stages of core collapse.  While this nucleus is not as massive as the nuclei that make up the bulk of the matter in this temperature-density regime ($A\sim 100$), it is similarly neutron rich and has allowed ground-state-to-ground-state isospin-raising charged current transitions with negative Q.  So, similar to the nuclei typical of this regime, it is in principle a good source of antineutrinos.  In this regime, however, positron capture and electron emission are suppressed by the high electron chemical potential and electron degeneracy, as there are few positrons to capture and little available outgoing electron phase space.  Antineutrino production through charged current processes is thus inhibited, and comparison with the deexcitation spectra in figure \ref{fig:nc_dec_spect} corroborates the suggestion by \cite{mbf:2013} that during stellar core collapse, neutral current deexcitation is likely the dominant source of antineutrinos.

Finally, two comments on nuclear structure issues are in order. First, the present study assumes exact isospin symmetry, but it is well known that this symmetry is only approximate.  There exists a few-percent discrepancy in several observed properties of pairs of mirror nuclei: in their masses \citep{kaneko-etal:2013}, in low-lying spectra \citep{kaneko-etal:2012}, and in results of charge-exchange experiments \citep{perez-loureiro-etal:2016}. Systematic exploration of isospin-symmetry breaking effects on nuclear astrophysics is an interesting and important topic.  Second, by way of introducing the ideas herein, the present study is restricted to computing $sd$-shell nuclei using OXBASH. Shell model calculations using improved theories for the $sd$-shell and up to the $fp+g_{9/2}$-shell are now available \citep{kaneko-etal:2015} and ready to be applied to the current interest. For arbitrarily heavy mass regions and deformed nuclei, we refer to the method of \cite{gsc:2006} developed using a novel type of shell model: the projected shell model of \cite{hs:1995} (for a recent review, see \cite{sun:2016}). Work along these lines is in progress. 

In summary, we provide with this work neutrino and antineutrino spectra for 70 $sd$-shell nuclei on the Fuller-Fowler-Newman temperature-density grid.  While previous authors have constructed rate tables, no such tabulation has previously been made for spectra.  Of particular note, there are currently no tables of neutral current deexcitation rates, let alone spectra; our tables remedy this.  We hope that these neutrino spectra will prove useful to stellar modelers of many stripes.  The tabulated neutrino spectra will be available for download from the Joint Institute for Nuclear Astrophysics - Center for the Evolution of the Elements (JINA-CEE) website.

\section{Acknowledgments}

We gratefully thank Surja Ghorui for his input in outlining the nuclear structure calculations.  We also thank Kelly Patton and Cecilia Lunardini for feedback on choosing the bounds and resolution of the neutrino spectra.  As ever, we thank B. Alex Brown for his assistance with OXBASH.  This research was supported at Shanghai Jiao Tong University by the National Natural Science Foundation of China (No. 11575112), the National Key Program for S\&T Research and Development (No. 2016YFA0400501), and the 973 Program of China (No. 2013CB834401); and at University of California, San Diego by  National Science Foundation Grants No. PHY-1307372 and PHY-1614864.

\software{OXBASH \citep{oxbash},
Matplotlib \citep{matplotlib}, NumPy \citep{numpy}, SciPy \citep{scipy}}

\bibliography{../references}

\end{document}